%% file: GLtransit2026.tex
\documentclass[twocolumn,trackchanges,showchanges]{aastex701}

\usepackage{ulem}
\usepackage{multirow}

\graphicspath{{./}{figures/}}

\begin{document}

\title{Parameter estimation from the transit light curve including gravitational lensing effects}

\author[0009-0004-4904-5792]{Shinta Kasuya}
\affiliation{Physics Division, Faculty of Science, Kanagawa University, Kanagawa 221-8686, Japan}
\email[show]{kasuya@kanagawa-u.ac.jp} 

\author[0000-0002-4909-5763]{Akihiko Fukui}
\affiliation{Komaba Institute for Science, The University of Tokyo, 3-8-1 Komaba, Meguro, Tokyo 153-8902, Japan}
\affiliation{Instituto de Astrof\'{i}sica de Canarias (IAC), 38205 La Laguna, Tenerife, Spain}
\email{afukui@g.ecc.u-tokyo.ac.jp} 

\author{Ayana Suzuki}
\affiliation{Physics Division, Faculty of Science, Kanagawa University, Kanagawa 221-8686, Japan}
\email{r202370049cj@jindai.jp}

\author[0000-0001-7209-9204]{Naomi Tsuji}
\affiliation{Institute for Cosmic Ray Research, The University of Tokyo, Chiba 277-8582, Japan}
\email{ntsuji@icrr.u-tokyo.ac.jp} 

\begin{abstract}

We study transit light curves of exoplanets, incorporating gravitational lensing 
effects, and explore the possibility of estimating planetary masses using only the 
transit light curve. We analyze existing data for planets with relatively large 
orbital distances ($>1$~AU) from Kepler and TESS to evaluate how well their masses 
can be constrained. For six exoplanets, we derive 2-$\sigma$ upper limits of 
200--2000 Jupiter masses. We also construct mock data to assess the feasibility of 
determining masses for exoplanets with larger orbital radii than currently known. 
Our results show that a precision of $3\times 10^{-6}$ is required to recover the 
true masses of planets with 10 (5) Jupiter masses at orbital radii of $a=20$ (100)~AU. 
Furthermore, we demonstrate that 
neglecting gravitational effects can lead to underestimation of exoplanet radii, 
particularly for planets with large orbital separations. Using mock data with a 
precision of $2\times 10^{-4}$, assuming a 30 minutes cadence 
with Kepler for a $K_p=14$ mag star,
we find that the estimated radii can be underestimated by 1--20\% for 5 
and 10 Jupiter-mass planets with $a=$ 10--100~AU. These results highlight the 
importance of including microlensing effects when modeling transit light curves 
of wide-orbit planets.
\end{abstract}






\section{Introduction} 

The first exoplanet around solar-like stars was detected by 
\citet{1995Natur.378..355M}. Since then, more than 6000 
exoplanets have been confirmed.\footnote{
\url{https://exoplanetarchive.ipac.caltech.edu/}}
The largest fraction of them were found by the transit method.
One observes the drops of the brightness of the host star during 
the surrounding exoplanet crossing one's line of sight to the star. 
The transit light curve is mainly used to constrain the radius of
the exoplanet among other physical parameters.

On the other hand, the mass of the exoplanet is hardly obtained from 
the transit observation. It is mostly determined from the complementary 
observation of the radial velocity, as first demonstrated by 
\citet{1995Natur.378..355M}.

In this paper, we explore the possibility of determining the mass only 
from the transit light curve.\footnote{
The mass of the close-in planet could in principle be derived by 
measuring precisely the light curve affected by ellipsoidal variations 
and/or relativistic beaming. Transit-timing variations is another 
method to measure planetary masses from transit light curves alone, 
which is however only applicable to multi-planet systems.} 
Since the exoplanet has mass, there should be some influence of the 
gravitational lensing when the light from the host star
passes by it. This means that we need to consider both occultation and 
lensing effects simultaneously. Microlensing with occultation has been 
considered generally or in other contexts, such as in the white dwarf 
binary \citep{1973A&A....26..215M,1996ApJ...467..537B,
2001MNRAS.324..547M,2002ApJ...579..430A,2003ApJ...594..449A,
2002A&A...394..489B,2002A&A...382....6B,2003ApJ...584.1042S,
2009ApJ...695..200L,2014Sci...344..275K,2016ApJ...820...53H,2018AJ....155..144K,2019ApJ...881L...3M}.
\citet{2011MNRAS.411.1863K} considered both effects on the transit of the 
exoplanets, paying special attention to establishing the clues of the 
microlensing effects. Using the model of the transit light curves with 
the lensing effect based on \citet{2011MNRAS.411.1863K}, we fit the actual 
transit light curves obtained by Kepler and TESS to see how much 
one can derive or constrain the mass of the exoplanets.

In addition to the estimation of the planetary mass, we would like to stress 
that the radius of the planet would be underestimated without the 
gravitational lensing effects taken into account. This is because the 
drop of the light curve becomes shallower by the lensing effects.

The structure of this paper is as follows. In the next section, we introduce 
a light-curve model that incorporates gravitational lensing effects. 
Section \ref{sec:KT} describes our application of a seven-parameter Markov 
Chain Monte Carlo (MCMC) analysis to estimate the exoplanet mass using Kepler 
and TESS data. Due to the small orbital radius, the derived mass constraints 
remain relatively weak. We then generate mock data for larger orbital 
radii and fit the corresponding model light curves to evaluate the precision 
of the observation required to recover the true planetary masses in 
Section\,\ref{sec:Mock}. In Section\,\ref{sec:radius}, we demonstrate that one would 
underestimate the planet radii using the transit model
that neglects lensing effects for the mock data 
which are created by including the microlensing effects. Section\,\ref{sec:concl} 
is devoted to our conclusions.

\section{Transit with gravitational lensing} \label{sec:GLtransit}
\begin{figure}[ht!]
\plotone{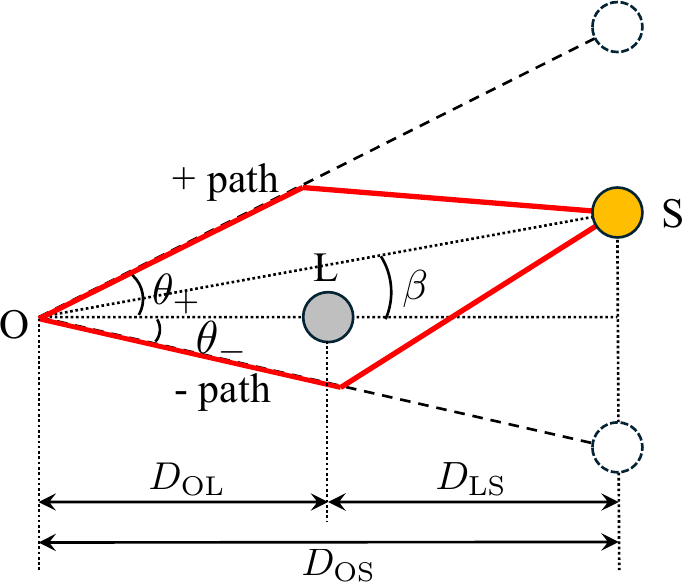}
\caption{Sketch of the geometry of the gravitational lensing. The light
from the star (S) is bent by the lens planet (L) and reaches to the
observer (O). There are two paths ($\pm$) of the light, which correspond 
to the solutions of the lens equation (\ref{lens-eq}).
\label{fig:lens}}
\end{figure}

The lens equation for a point mass is written as \citep{1986ApJ...304....1P}
\begin{equation}
\beta(\theta) = \theta - \frac{\theta_E^2}{\theta},
\label{lens-eq}
\end{equation}
where $\theta$ and $\beta$ are angular separations between the lens 
and the image of the source with and without lensing, respectively 
(see Figure~\ref{fig:lens}). $\theta_E$ is the Einstein angle,
\begin{equation}
\theta_E = \sqrt{\frac{4GM_L}{c^2}\frac{D_{\rm LS}}{D_{\rm OS}D_{\rm OL}}},
\label{E_radius}
\end{equation}
which represents an effective size of the lens. Here $D_{\rm OS}$ and 
$D_{\rm OL}$ are the distances from the observer to the source and the 
lens, respectively, while $D_{\rm LS}$ is the distance between the lens 
and the source. $M_L$ is the lens mass, $G$ is the gravitational 
constant and $c$ is the speed of light. Solving Equation(\ref{lens-eq}), 
we obtain the solutions
\begin{equation}
\theta_\pm = \frac{\beta\pm\sqrt{\beta^2+4\theta_E^2}}{2},
\end{equation}
where $\theta_{+(-)}$ is the angle of the image whose light comes 
from the far (near) side of the lens, which we will call $+ (-)$ 
path hereafter.

The magnification of the light of a point on the source star 
through the $\pm$ path is given by
\begin{equation}
A_\pm = \frac{d\theta_\pm \theta_\pm d\phi}{d\beta \beta d\phi} 
= \frac{1}{2} \pm \frac{\beta^2+2\theta_E^2}{2\beta\sqrt{\beta^2+4\theta_E^2}},
\end{equation}
due to the conservation of the surface brightness. In the case of 
microlensing, where the gravitational effect is relatively small such 
as in the case of exoplanets, one cannot discriminate these two magnified 
images separately, so the observed magnification should be the addition 
of the two as
\begin{equation}
A(\beta) = A_+ +|A_-| = \frac{\beta^2+2\theta_E^2}
{\beta\sqrt{\beta^2+4\theta_E^2}}.
\end{equation}
Since the source is usually an extended object, integrating over 
the source, we can obtain the total magnification as
\begin{equation}
{\cal A} = \frac{\rm area \ with \ lens}{\rm area \ without \ lens}
=\frac{\int A(\beta) \beta d\beta d\phi}{\int \beta d\beta d\phi}.
\label{mag}
\end{equation}

Now let us investigate the transit of the exoplanet where we need to 
consider the finite size effect of the lens planet. There are three 
cases for the light from a certain point on the source star: 
(a) both $+$ and $-$ paths are occulted, (b) only $-$ path is occulted, 
or (c) neither path is occulted. We must therefore use $A(\beta)$ 
itself for case (c), $A_+(\beta)$ for case (b) instead of $A(\beta)$ 
or do not integrate over the region where case (a) holds when we 
integrate over the source star in the numerator of Equation(\ref{mag}).

In addition, we must consider the limb darkening of the source star 
where the surface brightness of the star is not uniform, but has 
position dependence such that the center is brighter than the edge. 
The effect of limb darkening can be parameterize quadratically as
\begin{equation}
f_{LD}(\cos\mu) = 1- u_1 (1-\cos\mu) - u_2 (1-\cos\mu)^2,
\label{LD}
\end{equation}
where $\mu$ is an angle between the normal of the star surface and 
the line of sight, and $u_1$ and $u_2$ are constants of $O(0.1)$ 
\citep{1995A&AS..114..247C}. We need to calculate the light curve 
of the transit including limb darkening effects from Equation(\ref{mag}) 
with the limb darkening function (\ref{LD}) being inserted into the 
integrand in both the numerator and the denominator. 

\begin{figure}[ht!]
\plotone{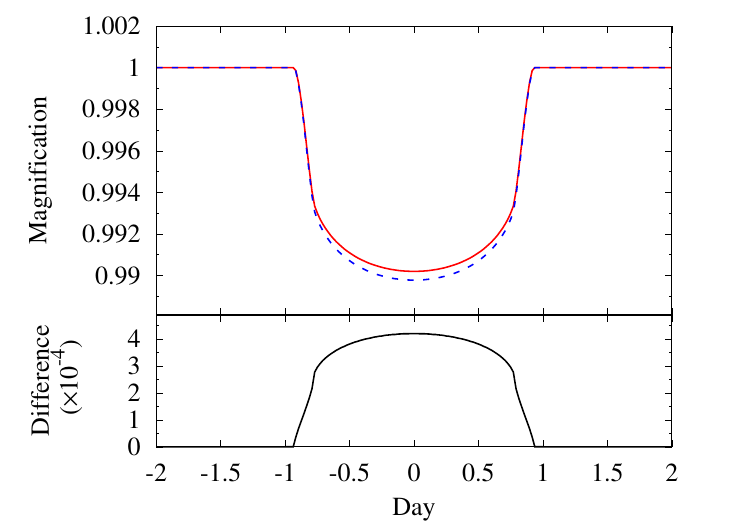}
\caption{Transit light curves with (red solid) and without (blue dashed) 
microlensing effects (top panel) and their difference (bottom panel). 
Here we adopt $M_L=10M_J$, $R_L=R_J$, and $D_{\rm LS}=10$~AU for the planet, 
while $M_S=M_\sun$, $R_S=R_\sun$, and $D_{\rm OS}=10$~pc for the source star.
\label{fig:lc_theory}}
\end{figure}

We show an example of theoretically calculated light curves with 
and without microlensing in Figure~\ref{fig:lc_theory}. Here we adopt  
$M_L=10M_J$, $R_L=R_J$, and $D_{\rm LS}=10$~AU, where $M_J$ and 
$R_J$ are the Jupiter mass and radius, respectively. For the source 
star, we adopt $M_S=M_\sun$, $R_S=R_\sun$, and $D_{\rm OS}=10$~pc. 
Notice that the dependence of $A(\beta)$ on $D_{\rm OS}$ is negligibly 
weak for $D_{\rm OS} \gg D_{\rm LS}$. At a glance, it is expected that 
one may underestimate the planet radius if one neglects the effects
of microlensing, where we can derive it roughly as 
$R_L=0.985 R_J$ from the amplitude at the transit center.

\begin{figure}[ht!]
\plotone{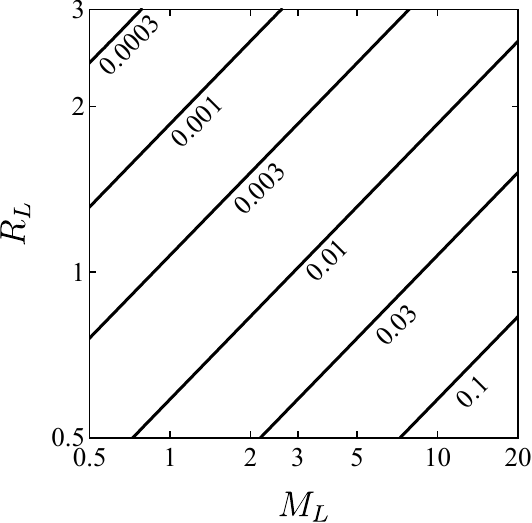}
\caption{Relative difference of the transit depths with and without  
microlensing effects, $(\Delta_{\rm NL}-\Delta_{\rm GL})/\Delta_{\rm NL}$, 
for $D_{\rm OS}=10$~pc, $D_{\rm LS}=10$~AU, $M_S=M_\sun$, and $R_S=R_\sun$.
\label{fig:depth}}
\end{figure}

The dependence of the lens mass and radius on the transit depth difference can be
easily seen if we consider the simple configuration that the centers of both lens 
planet and source star are aligned and limb darkening effects are neglected.
The transit depth is calculated as $\Delta_{\rm GL(NL)}=1-{\cal A}_{\rm GL(NL)}$,
where the subscript `GL (NL)' stands for the case with (without) 
lensing effects.
The magnification with and without lensing effects are given respectively by
\begin{eqnarray}
{\cal A}_{\rm GL} & = & 
\frac{1}{2}\left[ 1- \tilde{\beta}_L^2 + \sqrt{1+4\tilde{\theta}_E^2}
-\tilde{\beta}_L \sqrt{\tilde{\beta}_L^2+4\tilde{\theta}_E^2}\right], \\
{\cal A}_{\rm NL} & = & 1-\tilde{\theta}_L^2,
\end{eqnarray}
where $\theta_S=R_S/D_{\rm OS}$, $\theta_L=R_L/D_{\rm OL}$, $\beta_L=\beta(\theta_L)$,
and the tilde denotes the variables normalized with respect to $\theta_S$
\citep{2011MNRAS.411.1863K}.
We show the relative difference $(\Delta_{\rm NL}-\Delta_{\rm GL})/\Delta_{\rm NL}$ 
in the $(M_L, R_L)$-plane for $D_{\rm OS}=10$~pc, $D_{\rm LS}=10$~AU, $M_S=M_\sun$, 
and $R_S=R_\sun$ in Figure~\ref{fig:depth}. 

While recognizing the importance of accurately estimating the radius, 
we first focus on determining the mass and subsequently revisit the 
radius estimation in Section 5.

\section{Results of existing data by Kepler and TESS} \label{sec:KT}
We choose those planets that have relatively large orbital radii 
($> 1$~AU) among the Kepler and TESS samples, namely TOI-4600\,c, 
Kepler-1708\,b, Kepler-167\,e, Kepler-1654\,b, Kepler-421\,b, and 
Kepler-700\,c\footnote{
Although Kepler-700\,c is not yet confirmed as a planet, we include it in
our study, since Kepler-700\,b is confirmed as the inner planet.}. 
To analyze the light curves, we download time series 
photometry data of each object obtained with Kepler or TESS from 
the Mikulski Archive for Space Telescopes (MAST)\footnote{
https://doi.org/10.17909/3my2-5x97}. For all planets, two or 
more transit events are recorded in the respective data, of 
which we select one good transit event (with full transit coverage 
and minimal systematics) and extract a portion 
of the light curve around the transit for the subsequent analysis. 
For Kepler-1708\,b, Kepler-167\,e, Kepler-421\,b, and Kepler-700\,c, 
we use detrended and unwhitened light curves, a product called LC\_INIT, 
generated from the 29.4 minutes long-cadence data by the Data 
Validation (DV) module of the NASA Ames Science Operation Center 
pipeline \citep{2010ApJ...713L..87J,2018PASP..130f4502T}. For 
Kepler-1654\,b, for which no DV time series data are available, 
we download the Pre-search Data Conditioning Simple Aperture Photometry 
(PDCSAP) long-cadence light curve of Quarter 17. For TOI-4600\,c, 
we download the PDCSAP light curve produced from two-minute-cadence 
data of Sector 53, binning it by a factor of 5 
to save the computational time. A linear correction is applied to 
the out-of-transit baseline if a long-term trend is present in 
the light curve. 

Here, we conduct a seven-parameter MCMC analysis for these planets and 
investigate whether the planetary mass can be derived solely from the 
transit light curve. The seven parameters are the mass $M_L$ and the 
radius $R_L$ of the lens planet, the impact parameter $b$,  the time 
at the transit center $T_c$, the orbital radius $a$, and two limb 
darkening parameters $u_1$ and $u_2$. We assume circular orbits for 
simplicity. Note that no significant eccentricity ($>$0.4) has been reported for 
any of our sample. For the MCMC analysis, we use an affine-invariant 
ensemble sampler implemented in {\tt emcee} \citep{emcee}. 
Uniform priors are applied to all parameters except for $u_1$ and $u_2$, where  
Gaussian priors are adopted with mean values and standard deviations ($\sigma_{u_1}$ 
and $\sigma_{u_2}$) calculated using {\tt LDTK} \citep{2015MNRAS.453.3821P}. 
Note that, to take into account systematics in stellar models, we inflate 
the uncertainties estimated by {\tt LDTK} by a factor of 10 when deriving 
$\sigma_{u_1}$ and $\sigma_{u_2}$.

\begin{figure}[ht!]
\plotone{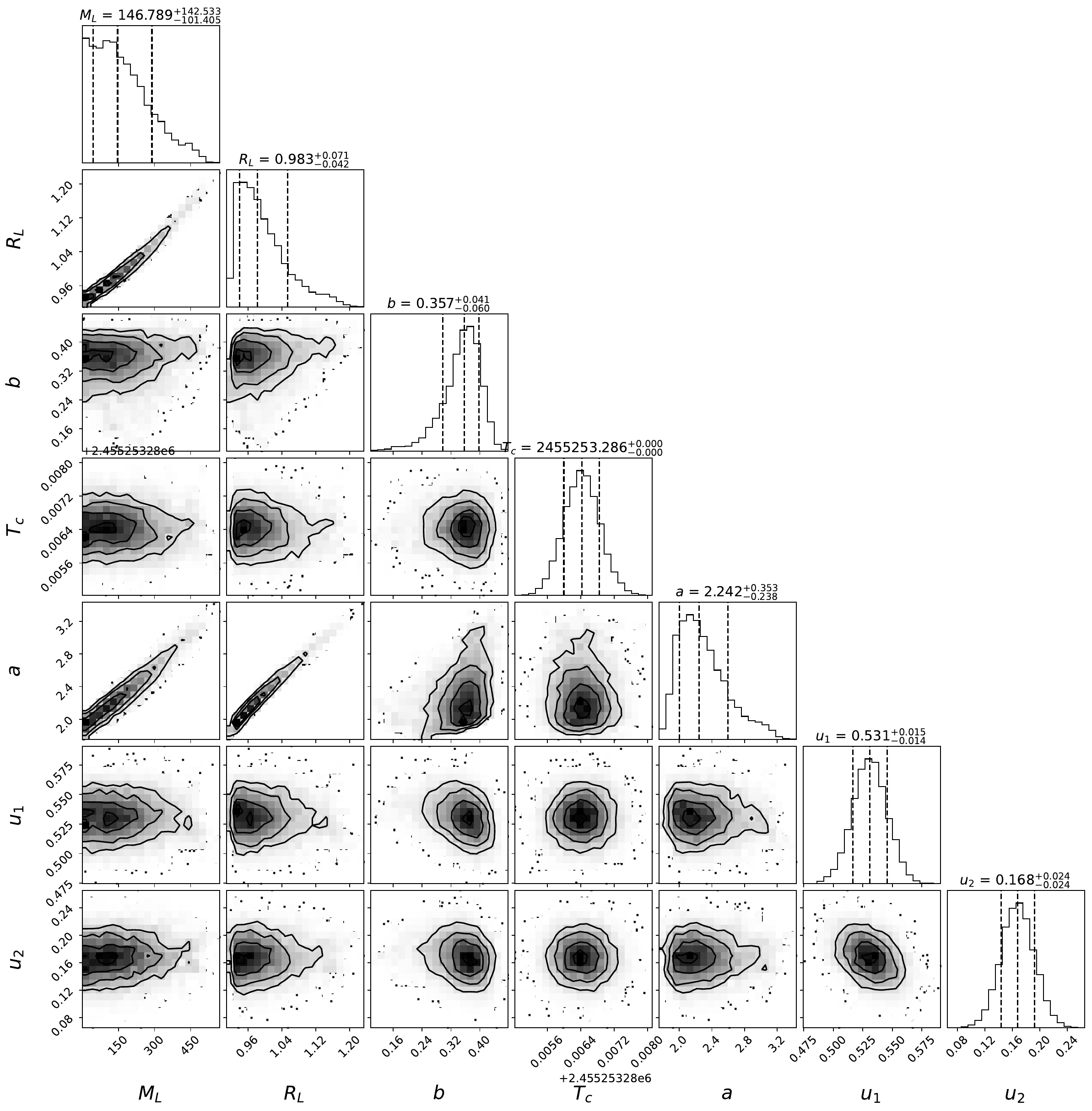}
\caption{Posterior distribution of full seven parameters for Kepler-167\,e.
\label{fig:corner-Kepler167e}}
\end{figure}

\begin{figure}[ht!]
\plotone{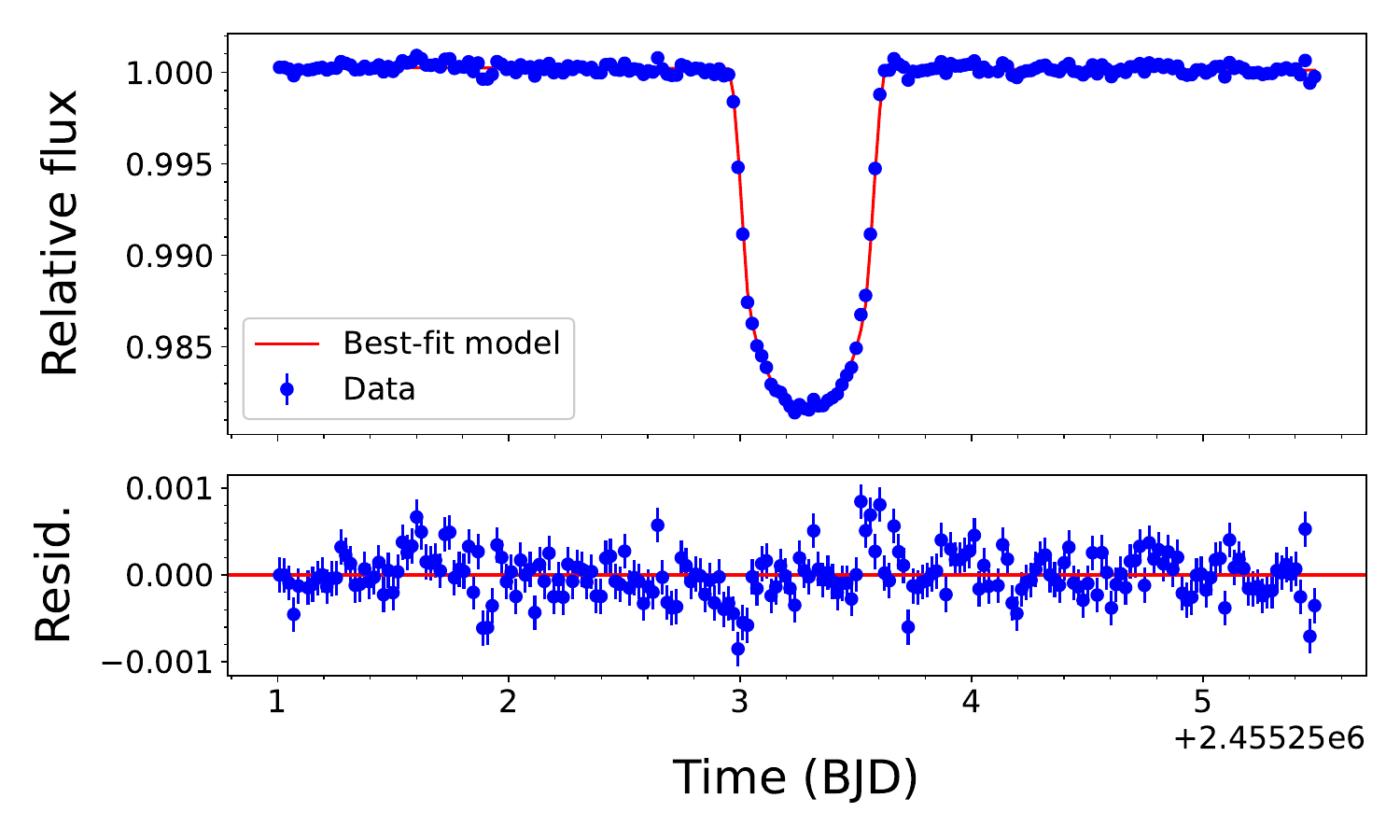}
\caption{Light curve of the best-fit model for Kepler-167\,e.
\label{fig:bestfit-Kepler167e}}
\end{figure}

We show posterior distributions of full seven parameters for Kepler-167\,e in 
Figure~\ref{fig:corner-Kepler167e}, and the light curve of the best-fit model in 
Figure~\ref{fig:bestfit-Kepler167e}. We can only constrain the exoplanet mass very 
loosely as $M_L < 396 M_J$ obtained by 2-$\sigma$ upper limit. As can be seen in 
the figure, there is degeneracy between $a$ and $R_L$, which can be understood 
that the magnification by the lensing effect for larger $a$ is compensated by more 
occultation by larger $R_L$. Another degeneracy between $M_L$ and $R_L$ arises
from the fact the amplification by the larger $M_L$ is offset by the reduction 
for larger $R_L$. We compile all the derived parameters in 
Table~\ref{tab:para_Kepler_TOI}.

\begin{table*}[ht!]
\begin{rotatetable*}
\centering
\caption{Evaluated values of parameters}\label{tab:para_Kepler_TOI}
\begin{tabular}{c|c|cccccc|c}
\hline\hline
Object & & $M_L~(M_J)$ & $R_L~(R_J)$ & $b$ & $a$~(AU) &
$u_1$\tablenotemark{a} & $u_2$\tablenotemark{a}  & \\
\hline
\multirow{2}{*}{Kepler-167e} 
& This work & $<396$ & $0.983^{+0.071}_{-0.042}$ & $0.375^{+0.041}_{-0.060}$ &
$2.24^{+0.35}_{-0.24}$ & $0.531^{+0.015}_{-0.014}$ & $0.168\pm 0.024$ & \\
& Literature & $1.01^{+0.16}_{-0.15} $ & $0.906\pm 0.038$ & $0.271^{+0.051}_{-0.073}$ &
$1.152\pm 0.068$ & $0.5062\pm 0.0019$\tablenotemark{a} & $0.1335\pm 0.0028$\tablenotemark{a} & 
\citet{2022ApJ...926...62C} \\
\hline
\multirow{2}{*}{TOI-4600c} 
& This work & $<2065$ & $1.39^{+0.21}_{-0.30}$ & $0.44^{+0.11}_{-0.18}$ &
$2.45^{+0.56}_{-0.75}$ & $0.466\pm 0.026$ & $0.133\pm 0.041$ & \\
& Literature & $<9.27$ & $0.840\pm 0.038$ & $0.46^{+0.10}_{-0.19}$ &
$1.152\pm 0.068$ & $0.4649\pm 0.0027$\tablenotemark{a} & $0.1461\pm 0.0044$\tablenotemark{a} & 
\citet{2023ApJ...954L..15M}\\
\hline
\multirow{2}{*}{Kepler-1708b} 
& This work & $<1035$ & $1.11^{+0.18}_{-0.12}$ & $0.21^{+0.18}_{-0.15}$ &
$2.36^{+0.53}_{-0.46}$ & $0.378^{+0.030}_{-0.029}$ & $0.165\pm 0.052$ & \\
& Literature & $<4.6$ & $0.889^{+0.054}_{-0.053}$ & $<0.37$ &
$1.64\pm 0.10$ & $0.3774\pm 0.0032$\tablenotemark{a} & $0.1637\pm 0.0055$\tablenotemark{a} & 
\citet{2022NatAs...6..367K} \\
\hline
\multirow{2}{*}{Kepler-1654b} 
& This work & $<883$ & $1.03^{+0.12}_{-0.16}$ & $<0.149$ &
$2.36^{+0.33}_{-0.48}$ & $0.419^{+0.016}_{-0.017}$ & $0.124^{+0.027}_{-0.025}$ & \\
& Literature & $<0.5$ & $0.819^{+0.019}_{-0.017}$ & $0.114^{+0.110}_{-0.079}$ &
$2.026^{+0.037}_{-0.035}$ & $0.4193\pm 0.0019$\tablenotemark{a} & $0.1514\pm0.0029$\tablenotemark{a} & 
\citet{2018AJ....155..158B} \\
\hline
\multirow{2}{*}{Kepler-421b} 
& This work & $<198$ & $0.471^{+0.065}_{-0.044}$ & $0.40^{+0.15}_{-0.23}$ &
$2.36^{+0.44}_{-0.32}$ & $0.463^{+0.029}_{-0.028}$ & $0.204^{+0.043}_{-0.039}$ & \\
& Literature & --- & $0.371^{+0.017}_{-0.014}$ & $0.21^{+0.17}_{-0.14}$ &
$1.219^{+0.089}_{-0.106}$ & $0.4373\pm 0.0031$\tablenotemark{a} & $0.1585\pm0.0047$\tablenotemark{a} & 
\citet{2014ApJ...795...25K} \\
\hline
\multirow{2}{*}{Kepler-700c} 
& This work & $<1236$ & $1.36^{+0.25}_{-0.32}$ & $0.830^{+0.024}_{-0.048}$ &
$4.0^{+1.1}_{-1.3}$ & $0.450^{+0.024}_{-0.023}$ & $0.149^{+0.037}_{-0.038}$ & \\
& Literature & --- & $0.51\pm 0.22$ & $0.776^{+0.048}_{-0.013}$ &
$1.24\pm 0.25$\tablenotemark{b} & $0.4475\pm 0.0024$\tablenotemark{a} & $0.1384\pm 0.0039$\tablenotemark{a}
& \citet{2016ApJ...822....2U} \\
\hline
\end{tabular}
\tablenotetext{\rm a}{Calculated by using {\tt LDTK} \citep{2015MNRAS.453.3821P}.}
\tablenotetext{\rm b}{Calculated by using values of $a/R_*$ and $R_*$ in \citet{2015ApJ...815..127W}.}
\end{rotatetable*}
\end{table*}

We apply the same procedure to other exoplanets: TOI-4600\,c, Kepler-1708\.b, 
Kepler-1654\,b, Kepler-421\,b, and Kepler-700\,c. The summary is shown in 
Table~\ref{tab:para_Kepler_TOI}, and all the posterior distributions are displayed in 
Figures~\ref{fig:corner_Kepler167e_TOI4600c}--\ref{fig:corner_Kepler421b_Kepler700c} 
in Appendix~\ref{app:KT}. 

As shown in Table~\ref{tab:para_Kepler_TOI}, the masses of the first four planets are 
constrained by other methods. However, we can only derive very 
loose upper bounds on their masses: $396M_J$ for Kepler-167\,e, $2065M_J$ for
TOI-4600\,c, $1035M_J$ for Kepler-1708\,b, and $876M_J$ for Kepler-1654\,b.
On the other hand, since the masses of the latter two planets are not known, 
we were able to place direct upper limits on their masses for the first time: 
$198M_J$ for Kepler-421\,b and $1236M_J$ for Kepler-700\,c. However, we emphasize 
that these constraints are substantially weaker than the upper limits inferred 
from stellar models combined with plausible planetary density assumptions.

\section{Prospects for determining the planetary masses for larger orbital radii}
\label{sec:Mock}
Since the orbital radii are small for the existing data, we could not constrain 
the masses of the exoplanets tightly, as seen in the previous section. Then let 
us now generate mock data for larger orbital radii where the gravitational 
lensing effects would have relatively greater impacts on the light curves. Here 
we fit the corresponding model light curve to evaluate the precision required 
to recover the true planetary mass. 

In order to obtain the mock light curve data, we fix the mass and radius of the 
hosting star as $M_S=M_\sun$ and $R_S=R_\sun$, while adopting the values of the 
quadratic limb darkening parameters as $u_1=0.433$ and $u_2=0.156$, calculated 
by using {\tt LDTK} for the Sun-like star. The distance to the star is set to 10~pc, 
since its dependence on the light curve is negligible. As for the planet, we 
assume the radius as $R_L=R_J$, and the impact parameter as $b=0$, and vary 
the orbital radius adopting the values of $a=10$, 20, 100, and 200 AU (100 and 200~AU) 
for $M_L=10 M_J$ ($5 M_J$).
The mock light curves are calculated with a cadence of about 30 minutes for the cases 
of 10 and 20 AU, and about 120 minutes for the cases of 100 and 200 AU. The four times 
longer cadence for the 100 and 200 AU cases assumes a mosaic survey of four fields, 
motivated by the fact that the number of stars in the single Kepler field ($\sim10^5$) 
is too small to detect a transit of such large-orbit planets \citep{2011MNRAS.411.1863K}. 
The model fluxes are scattered with gaussian distribution of order of some small number, 
which represent the precision, together with error bars of order of the same precision. 
The precisions that we test with are $1\times10^{-6}$, $3\times10^{-6}$, $1\times10^{-5}$, 
$3\times10^{-5}$, and $1\times10^{-4}$.

The results are summarized in Table~\ref{tab:preci_mock2}. Here we show the 
precision which we could correctly estimate the mass of the exoplanet in each 
case, where the posterior distributions are displayed in  
Figures~\ref{fig:mock_10_10-20}--\ref{fig:mock_5_100-200} in Appendix~\ref{app:Mock_M}. 
We find that it is possible for the mission of the photometric precision as high as 
the Kepler \citep[$\sim10^{-4}$ per 30 minutes for a $K_p=12$ mag star;][]{2010ApJ...713L.120J} 
to estimate the mass of the exoplanet with $10M_J$ for $a=200$~AU. 
Moreover, one can derive the planetary mass 
$10M_J$ ($5M_J$) correctly for $a=20$~AU ($a=100$~AU) with higher-precision 
photometry of $3\times 10^{-6}$ in a bit far future as the near-future missions
such as PLATO \citep{2025ExA....59...26R} and Earth 2.0 \citep{2022arXiv220606693G} 
are not expected to reach such high precision.

\begin{table}[ht!]
\caption{Precision needed for $M_L=10 M_J$ ($a=10$ to 200~AU), and $M_L=5 M_J$ 
($a=100$ and 200~AU).}\label{tab:preci_mock2}
\centering
\begin{tabular}{c|c|c}
\hline\hline
(AU) & $M_L=5 M_J$ & $M_L=10 M_J$  \\
\hline
$a=10$   & --- & $1\times 10^{-6}$ \\
$a=20$   & --- & $3\times 10^{-6}$ \\
$a=100$ & $3\times 10^{-6}$ & $3\times 10^{-5}$ \\
$a=200$ & $1\times 10^{-5}$ & $1\times 10^{-4}$ \\
\hline
\end{tabular}
\end{table}

The observability of this kind of the transit could 
be estimated optimistically as 
\begin{eqnarray}
& & \hspace{-4mm}
P  \sim \frac{t_{\rm obs}}{T_{\rm orbit}}\times \frac{R_S}{a} \times N_S 
\nonumber \\
& & \hspace{-3.5mm}
\sim 1
\left(\frac{t_{\rm obs}}{4~{\rm yr}}\right)
\left(\frac{T_{\rm orbit}}{10^2~{\rm yr}}\right)^{-1}
\left(\frac{a}{20~{\rm AU}}\right)^{-1}
\left(\frac{R_S}{R_\odot}\right)
\left(\frac{N_S}{10^5}\right),
\nonumber \\
& & \hspace{-3.5mm}
\sim 1
\left(\frac{t_{\rm obs}}{10~{\rm yr}}\right)
\left(\frac{T_{\rm orbit}}{10^3~{\rm yr}}\right)^{-1}
\left(\frac{a}{100~{\rm AU}}\right)^{-1}
\nonumber \\
& & \hspace{30mm}\times \left(\frac{R_S}{R_\odot}\right)
\left(\frac{N_S}{2\times 10^6}\right),
\end{eqnarray}
where $t_{\rm obs}$ is operation time of observation mission, $N_S$ is the number
of the stars which will be searched in the mission, $a$ and $T_{\rm orbit}$ are 
respectively the orbital radius and period, and $R_S$ is the radius of the host star. 
The first and second factors in the first line represent the chance to meet the transit,
and the probability of the edge-on to the line of the sight.
Therefore, we may expect to find one exoplanet with a separation of 20~AU (100~AU) 
by those missions that search $10^5$ ($2\times 10^6$) stars in 4 (10) years.
Although detecting planets at 100 AU remains challenging, planets at 20 AU are 
likely to be within the reach of past, current, and planned missions such as Kepler, 
TESS, PLATO, and Earth 2.0.

\section{Misestimation of planet radii} \label{sec:radius}
In this section, we investigate how one could misestimate planet radii if one 
does not take lensing effects into account. To this end, assuming a 30 minutes cadence, 
we make mock light curve data for $M_L=5$ and $10 M_J$ with $a=10$, 20, and 100~AU, 
including the effects of microlensing by the planet. Here we fix $R_L=R_J$, 
$R_S=R_\sun$, $M_S=M_\sun$, and $d=10$~pc. The precision is taken to be 
$2\times 10^{-4}$, which corresponds to a precision of 200~ppm per 30 minutes 
cadence with Kepler for a $K_p=14$~mag star \citep{2010ApJ...713L.120J}.
We conduct MCMC analysis for the usual model without lensing effects using 
the Quadratic Model of {\tt PyTransit} \citep{Parviainen2015}\footnote{https://pytransit.readthedocs.io/}.
The summary is presented in Table~\ref{tab:radius}, and all the posterior
distributions are displayed in Figures~\ref{fig:mock_5_10_10}--\ref{fig:mock_5_10_100}
in Appendix~\ref{app:Mock_R}. Here we can see that one 
would underestimate the radius of the planet, especially for those cases 
with larger orbital radius and larger planetary mass: 1--20\% less than the
true values for these adopted parameters. 
It should be noted that the degree of misestimation of the planetary radius 
in the case $M_L=10 M_J$ and $a=10$~AU is consistent with the prediction in
Sect.~\ref{sec:GLtransit}.

\begin{table}[ht!]
\caption{Planetary radius estimation neglecting 
gravitational lensing effects}\label{tab:radius}
\centering
\begin{tabular}{cc|c}
\hline\hline
$M_L$ & $a$ & $R_L$   \\
$(M_J)$ & (AU) & $(R_J)$ \\
\hline
5 &  10 & $0.991^{+0.006}_{-0.005}$ \\
5 &  20 & $0.986\pm 0.003$ \\
5 & 100 & $0.910\pm 0.002$ \\ 
\hline
10 &  10 & $0.988\pm 0.005$ \\
10 &  20 & $0.969\pm 0.003$ \\
10 & 100 & $0.813\pm 0.002$ \\ 
\hline
\end{tabular}
\end{table}

Although no transiting planets with such large orbital separations have 
yet been confirmed, single-transit planetary candidates have been identified 
in Kepler and TESS survey data \citep[e.g.,][]{2019AJ....157..218K,2024MNRAS.528.2851M}, 
some of which may correspond to such systems. It is anticipated that the masses and 
orbits of these planetary candidates will be measurable using the astrometric method 
with time-series data from Gaia DR4 and subsequent releases 
\citep{2014ApJ...797...14P,2025MNRAS.536.2485W}. In such cases, it is not 
necessary to constrain the planetary masses from the transit light curves,
which is difficult with the photometric precision of Kepler, as shown in 
Sect.\,\ref{sec:Mock}. However, the microlensing effects may still affect the 
estimates of the planetary radii, and it will therefore be important to include 
this effect when modeling the light curves.

\section{Conclusions} \label{sec:concl}
We have studied the transit light curves of exoplanets taking into account 
their gravitational lensing effects, and explored the possibility of obtaining 
the masses of exoplanets only by using the light curves of the transits. To this end,
we must consider both occultation and lensing effects simultaneously. 
We have investigated the existing data from Kepler and TESS whether or not we 
can derive the masses which are determined from other methods such as the radial 
velocity, and found that we could have obtained only very loose upper limits,
since the orbital separations are not so large ($\lesssim 2$~AU) for those planets
observed by Kepler and TESS. The derived upper bounds are $396M_J$ for 
Kepler-167\,e, $2065M_J$ for TOI-4600\,c, $1035M_J$ for Kepler-1708\,b, 
$876M_J$ for Kepler-1654\,b, $198M_J$ for Kepler-421\,b and $1236M_J$ for 
Kepler-700\,c. Notice that we have placed the upper limits of the last two 
for the first time, although very loosely.

Moreover, for future prospects, we have constructed the mock data to find 
the ability to extract the exoplanet masses for larger orbital radii than 
those of existing data. This should be important because we can observe the 
transit only once for those exoplanets with such large orbital radii and 
radial velocity may be hard to be measured as well. We have obtained the 
precision that is necessary to retrieve the true values of the planetary masses. 
We have found that, in the case with $M_L=10 M_J$ ($5M_J$) and 
$a=20$~AU (100~AU), the correct masses could be derived by those future 
missions that have precision of $3\times 10^{-6}$, which is ten times better 
than the present best precision and may be achieved in the future.

In addition, we have pointed out that one would underestimate the radii
of exoplanets especially for large orbital radii if one neglects gravitational 
effects on the transit light curves. We have made mock light curve data 
including the effects of microlensing by the planet, which have precision 
of $2\times 10^{-4}$, assuming 30 minutes cadence for $K_p=14$~mag for Kelper.
The estimation of planet radii could be 1-20\% less than the true values for
the parameters that we have taken. Our results should suggest that it would 
be important to take microlensing effects into account when modeling 
the transit light curves.

\section*{Acknowledgment} 

We thank the anonymous referee for the helpful comments.
A.F. acknowledges support through JSPS KAKENHI Grant Numbers JP24K00689 and JP24H00017.

This paper includes data collected by the Kepler and TESS missions and obtained 
from the MAST data archive at the Space Telescope Science Institute (STScI). 
Funding for the Kepler and TESS missions are provided by the NASA Science Mission 
Directorate. STScI is operated by the Association of Universities for Research 
in Astronomy, Inc., under NASA contract NAS 5201326555. 

\newpage
\input{GLtr_Appendix}


\bibliography{GLtransit2026}{}
\bibliographystyle{aasjournalv7}


\end{document}

%% file: GLtr_Appendix.tex
\appendix
\section{Posterior distributions of parameters for the Kepler and TESS data} 
\label{app:KT}
\begin{figure}[hb!]
\plottwo{GLtr_Kepler-167e_tra1_corner2}{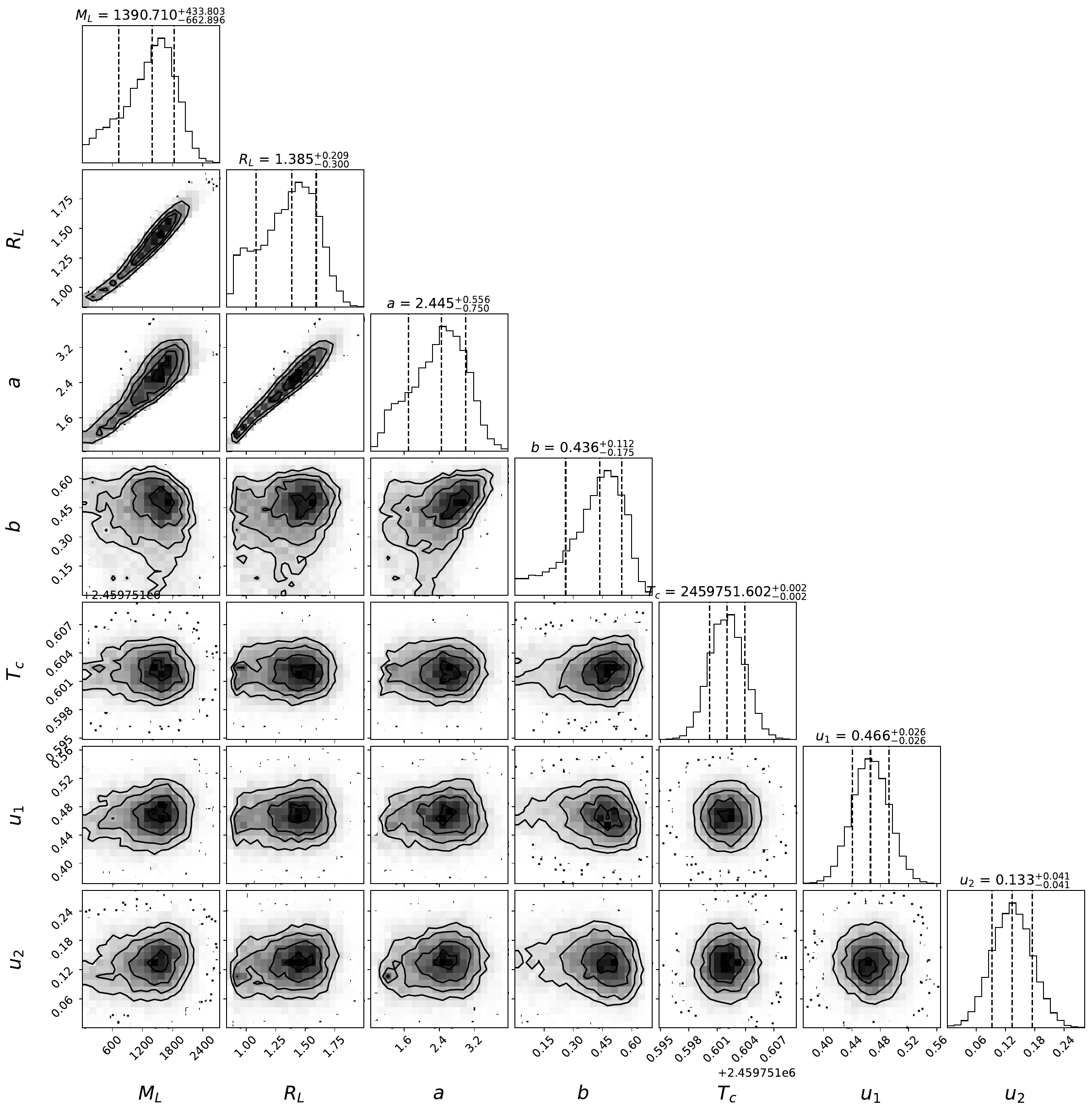}
\caption{Posterior distributions of full seven parameters for Kepler-167\,e (left) and 
TOI-4600\,c (right).
\label{fig:corner_Kepler167e_TOI4600c}}
\end{figure}

\begin{figure}[ht!]
\plottwo{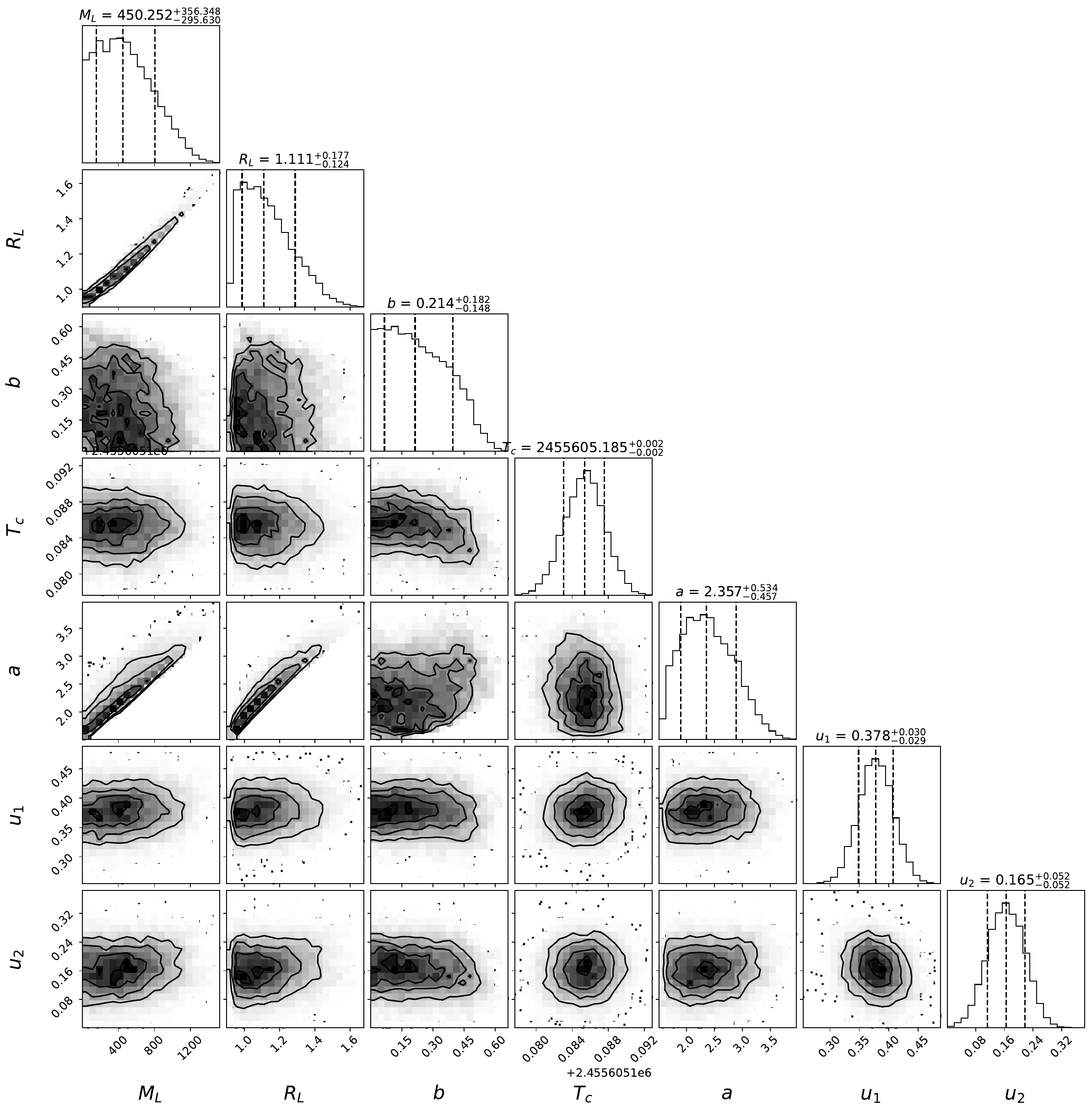}{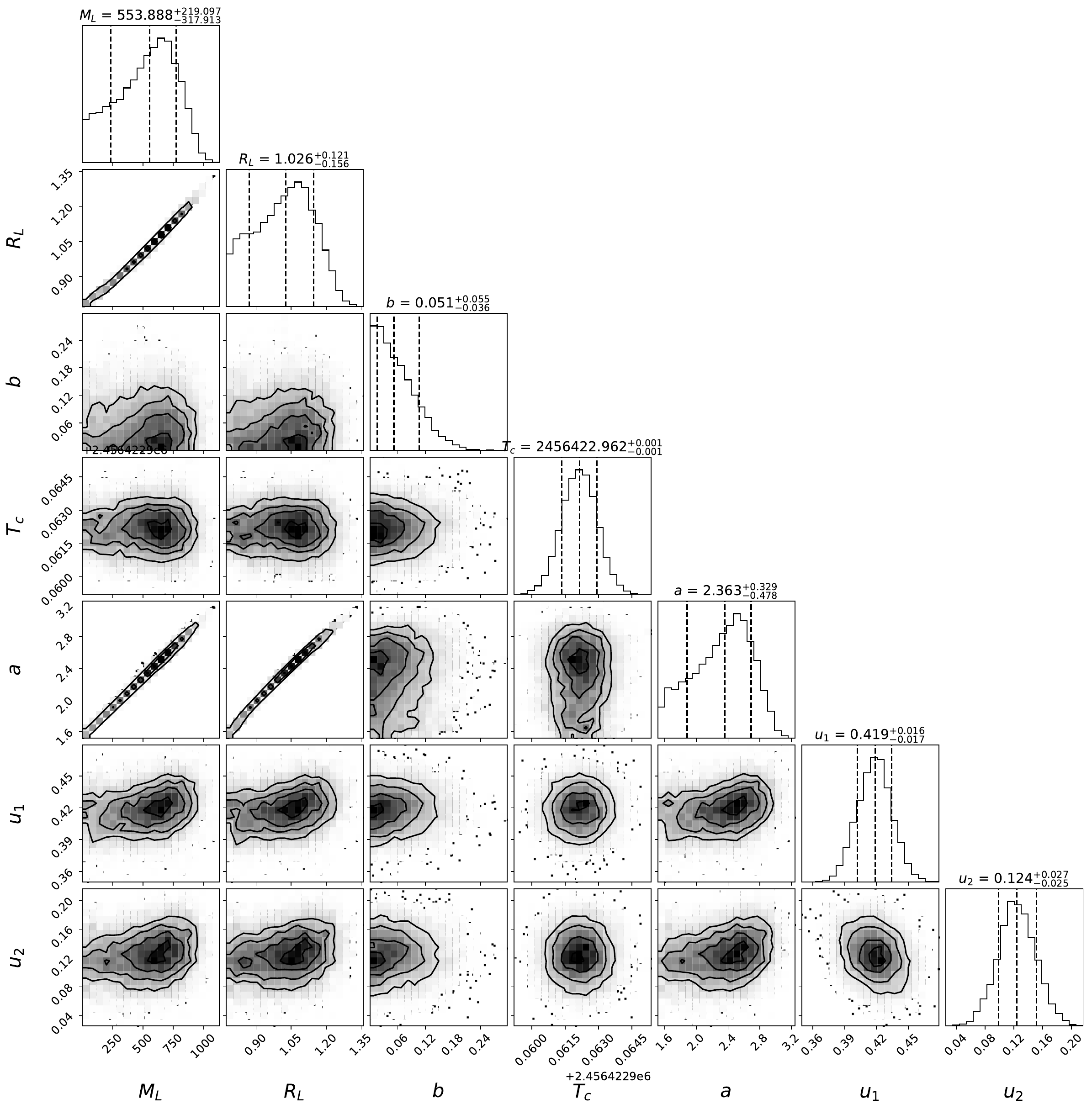}
\caption{Posterior distributions of full seven parameters for Kelper-1708\,b (left) and 
Kepler-1654\,b (right).
\label{fig:corner_Kelpler1708b_Kepler1654b}}
\end{figure}

\begin{figure}[ht!]
\plottwo{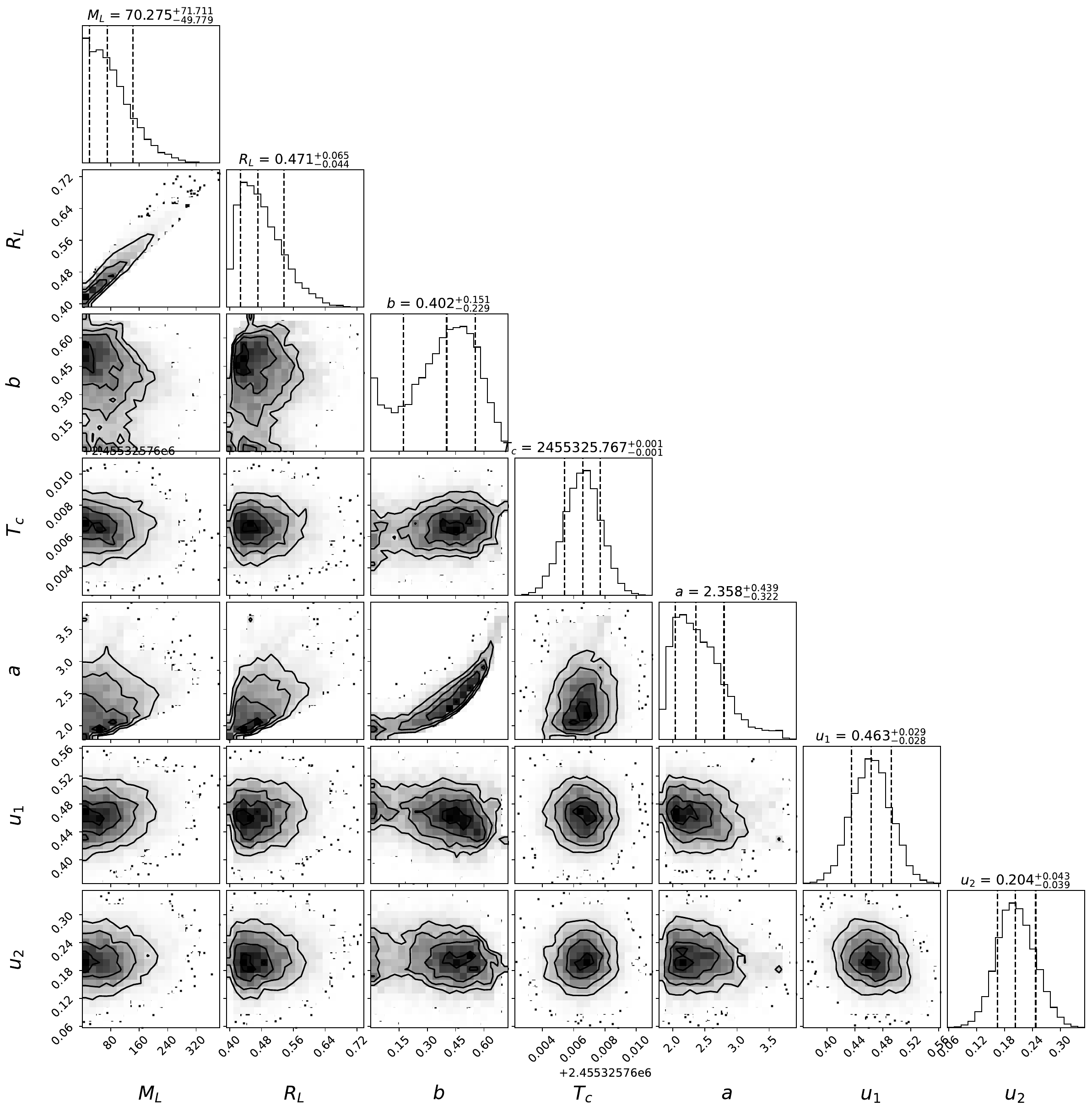}{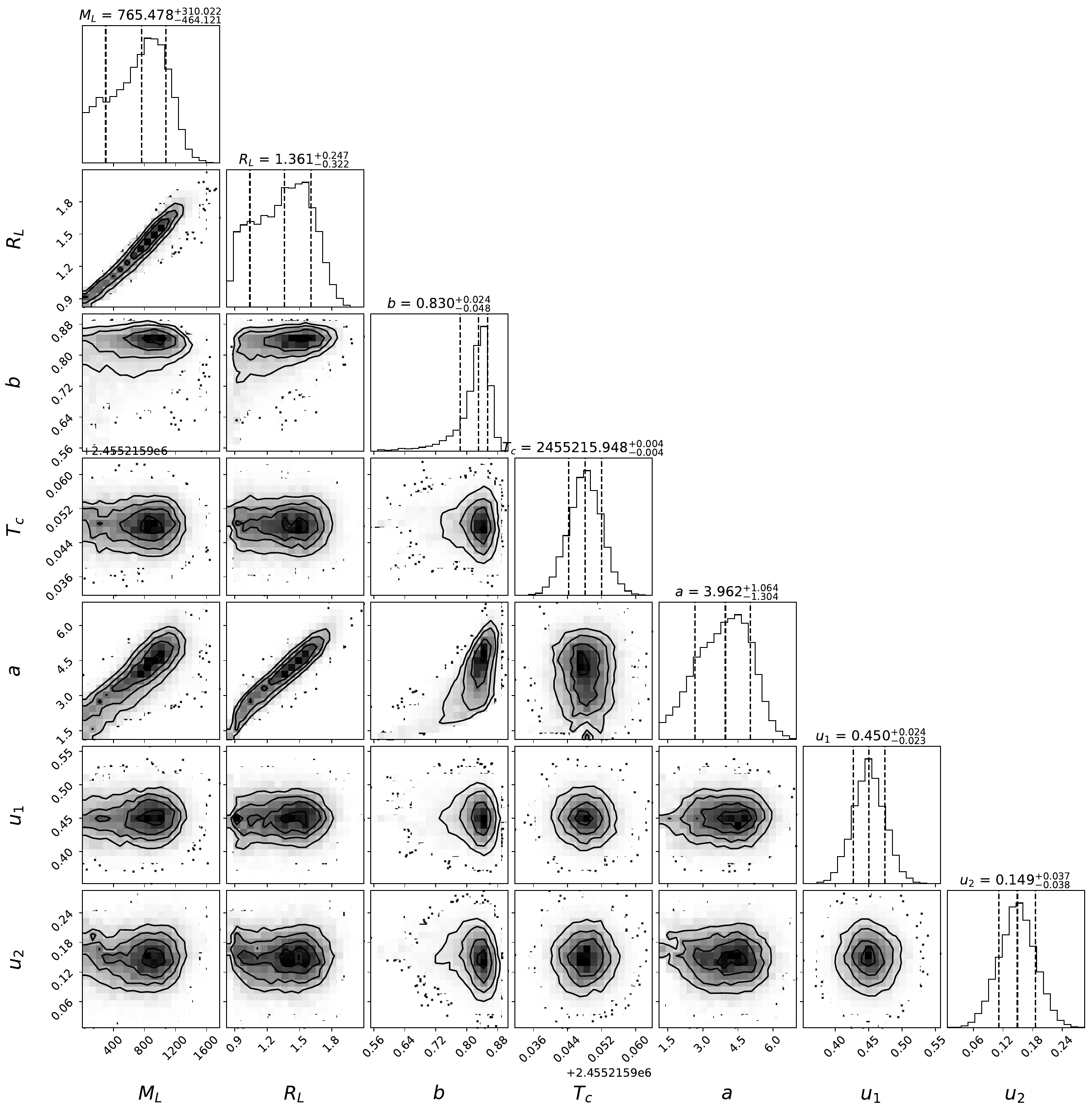}
\caption{Posterior distributions of full seven parameters for Kepler-421\,b (left) 
and Kepler-700\,c (right).
\label{fig:corner_Kepler421b_Kepler700c}}
\end{figure}

We show below the posterior distributions of full parameters for all the cases 
in the main text for the Kepler and TESS data in Figures \ref{fig:corner_Kepler167e_TOI4600c}-\ref{fig:corner_Kepler421b_Kepler700c}.

\vspace*{5mm}
\vspace*{5mm}

\vspace*{5mm}
\section{Posterior distributions of parameters for the mock data in Sect.~4}
\label{app:Mock_M}
\begin{figure}[ht!]
\plottwo{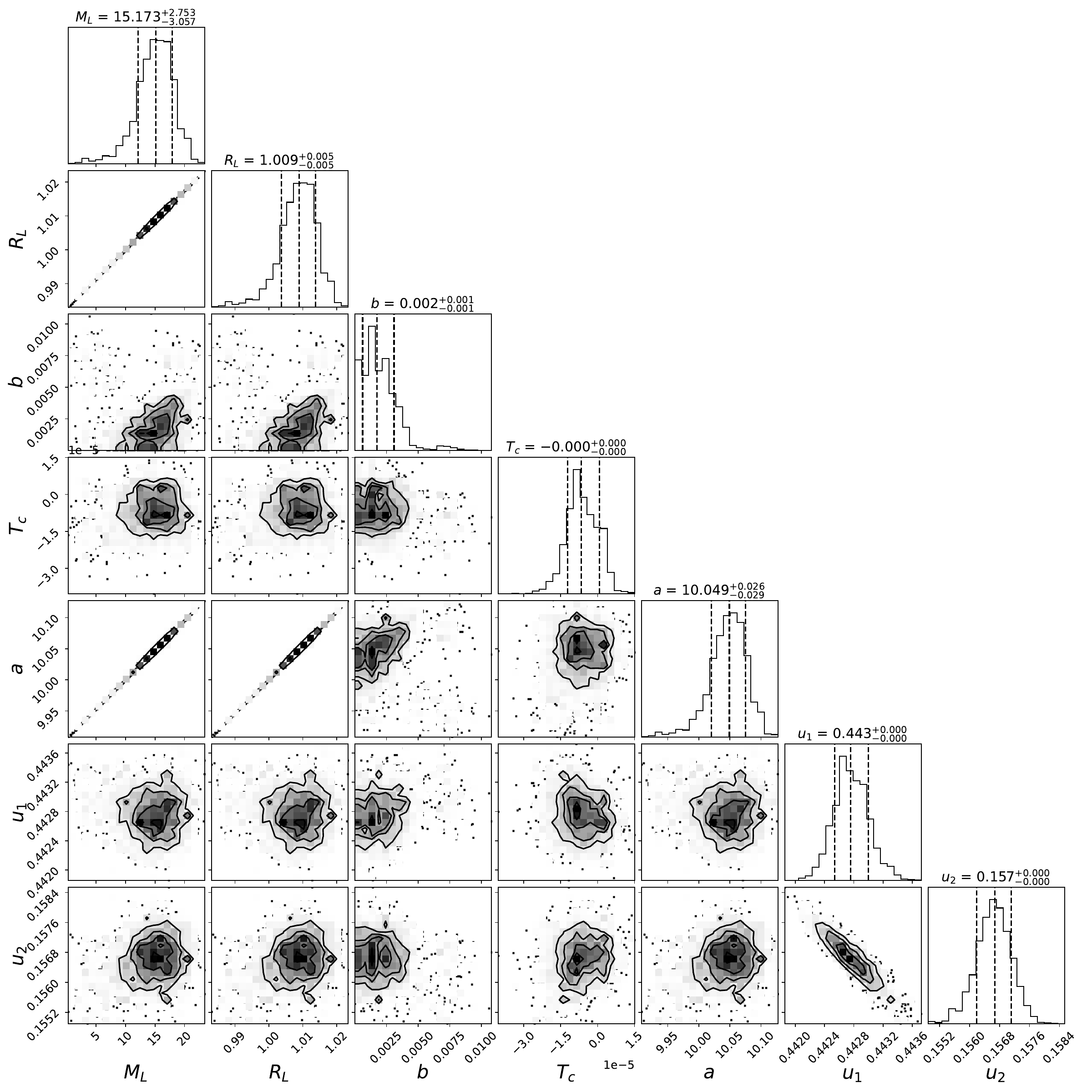}{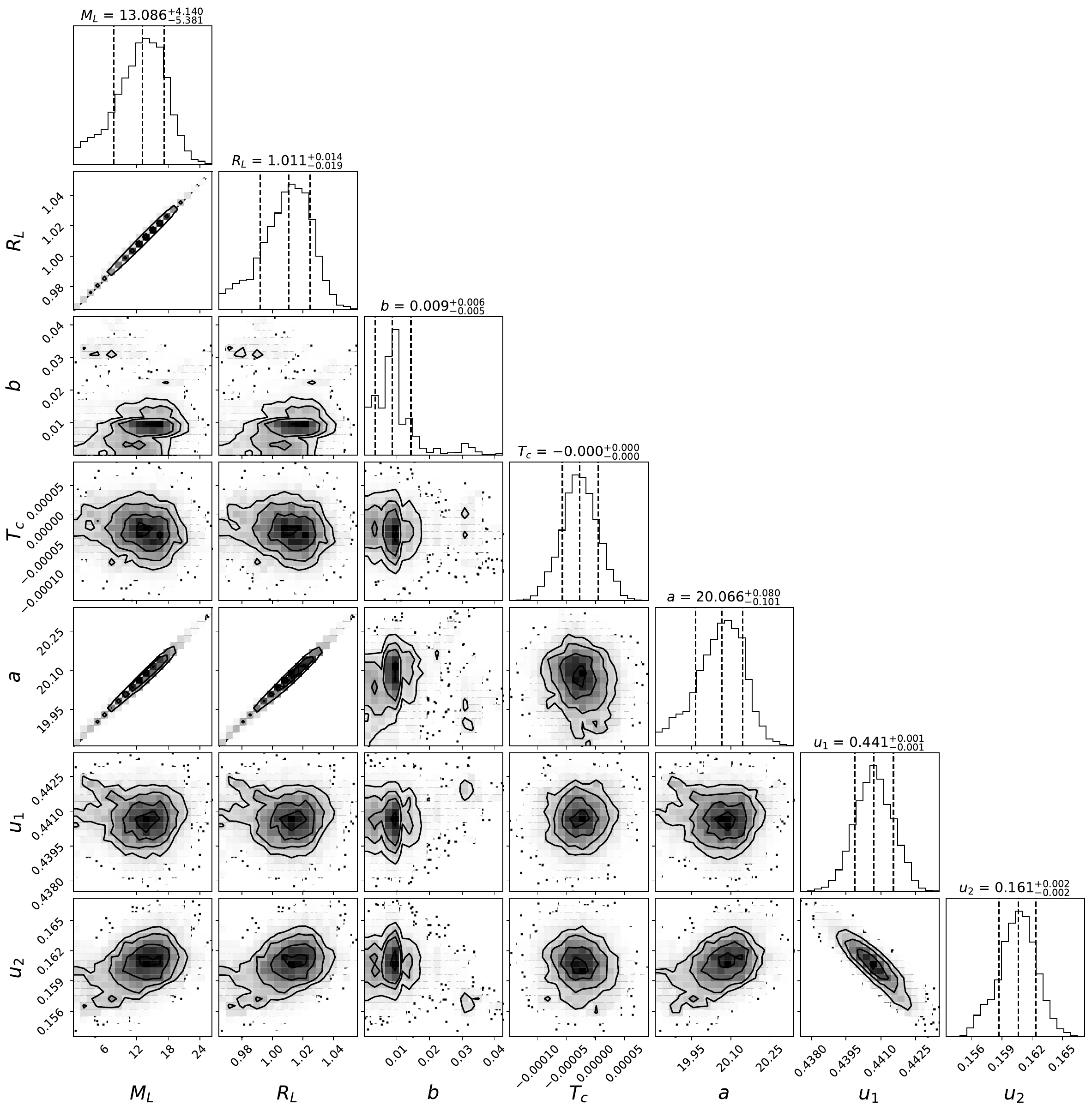}
\caption{Posterior distribution for the mock data with $a=10$~AU (left) and 
$a=20$~AU (right) for $M_L=10M_J$ including lensing effects.
\label{fig:mock_10_10-20}}
\end{figure}

\begin{figure}[ht!]
\plottwo{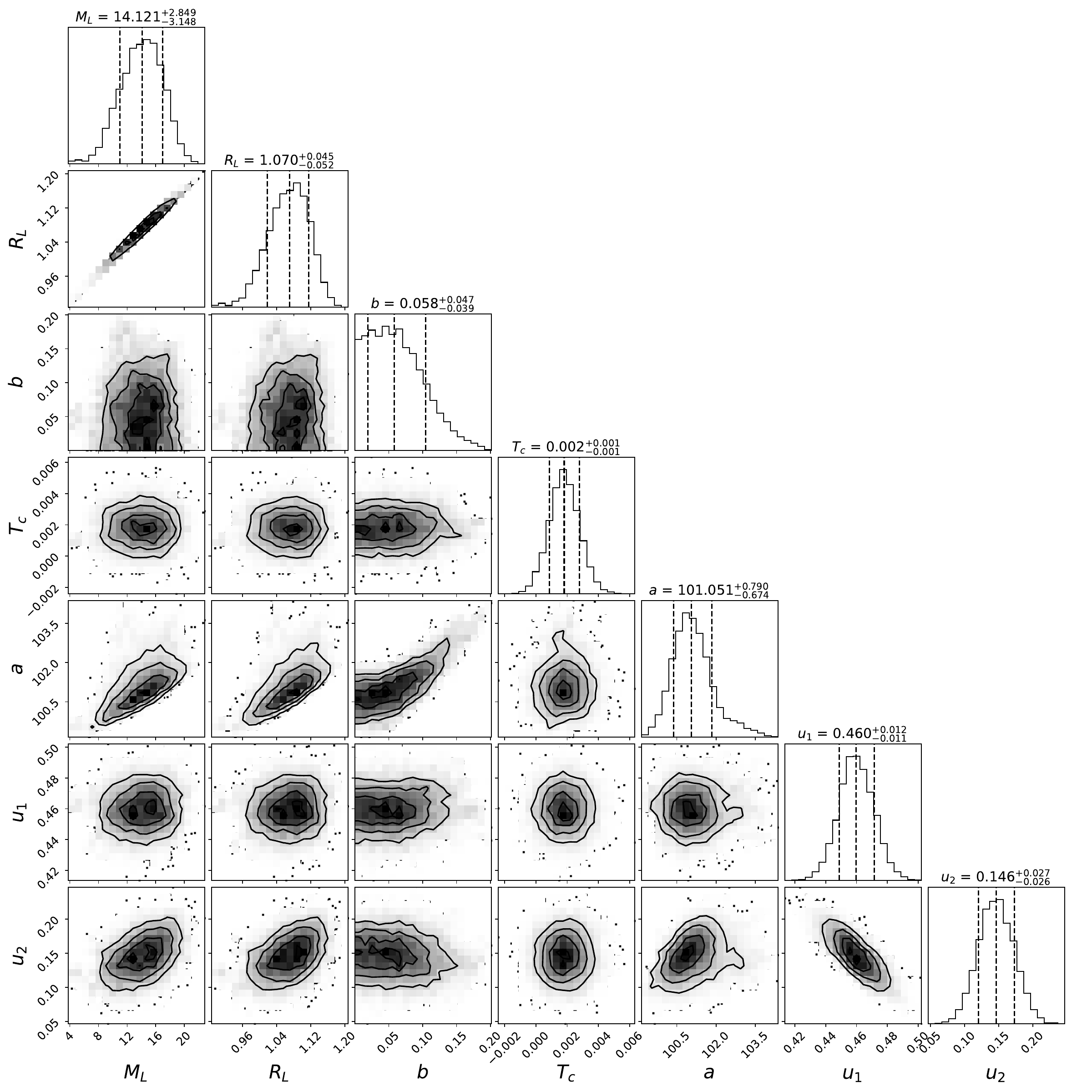}{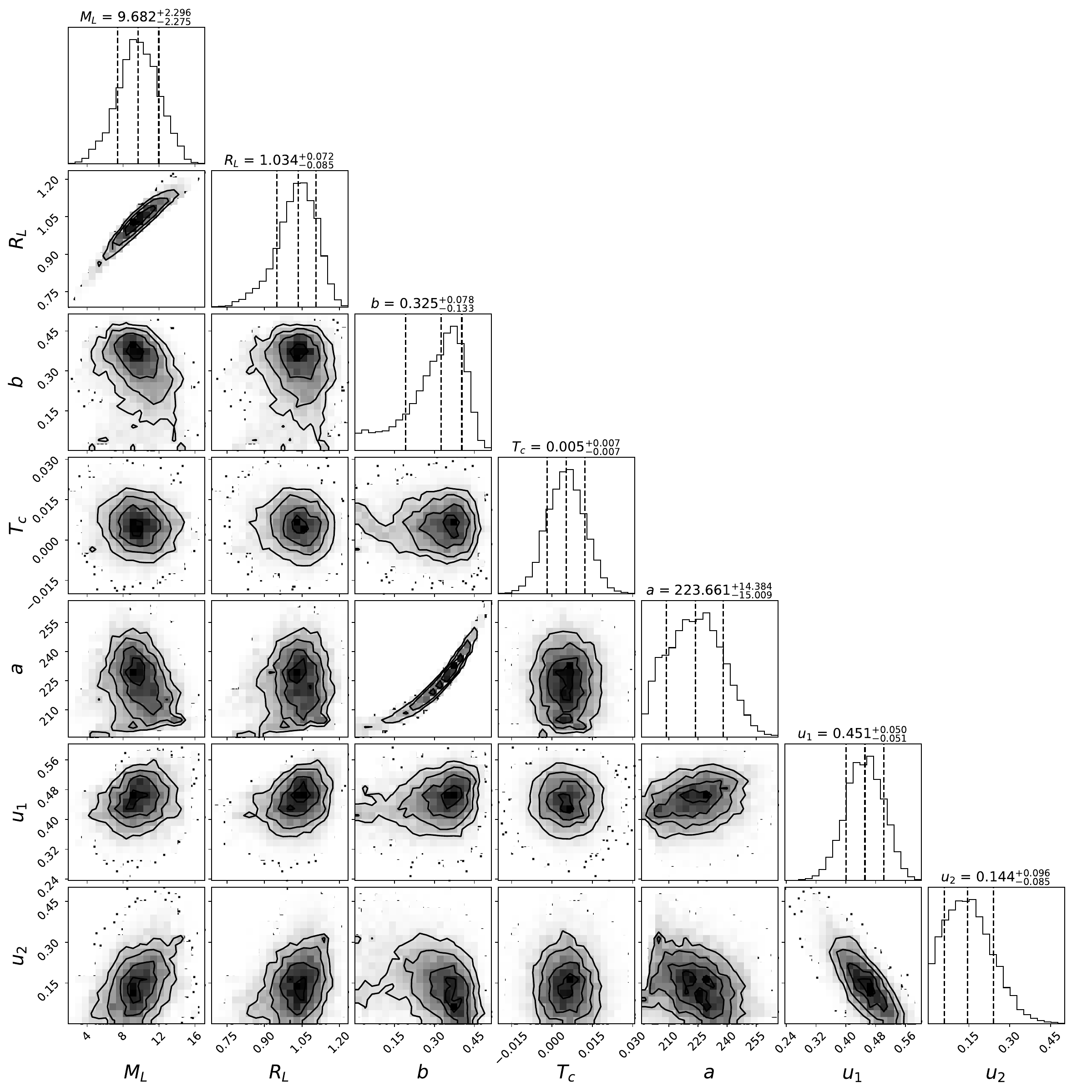}
\caption{Posterior distribution for the mock data with $a=100$~AU (left) and 
$a=200$~AU (right) for $M_L=10M_J$ including lensing effects.
\label{fig:mock_10_100-200}}
\end{figure}

\begin{figure}[ht!]
\plottwo{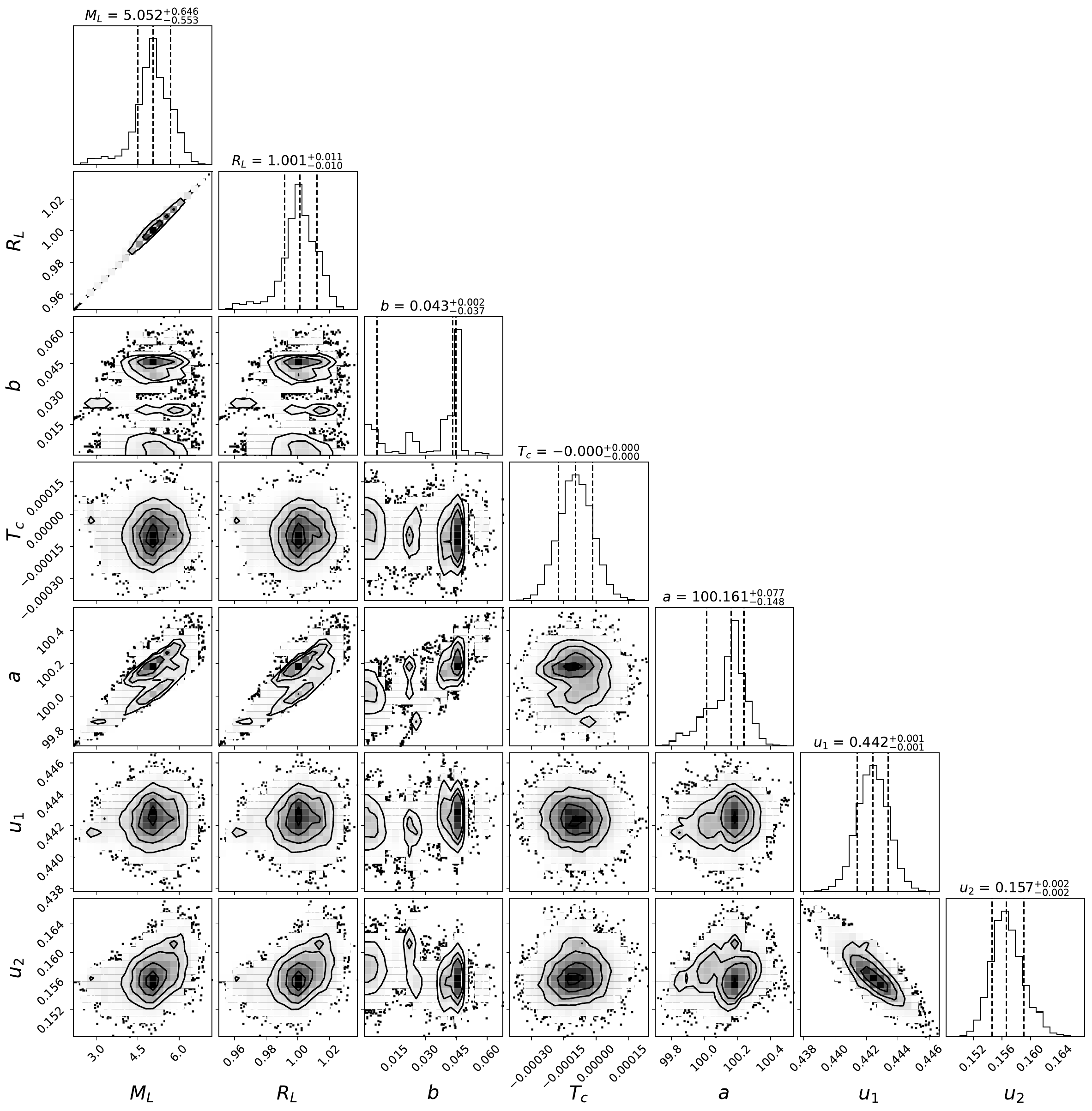}{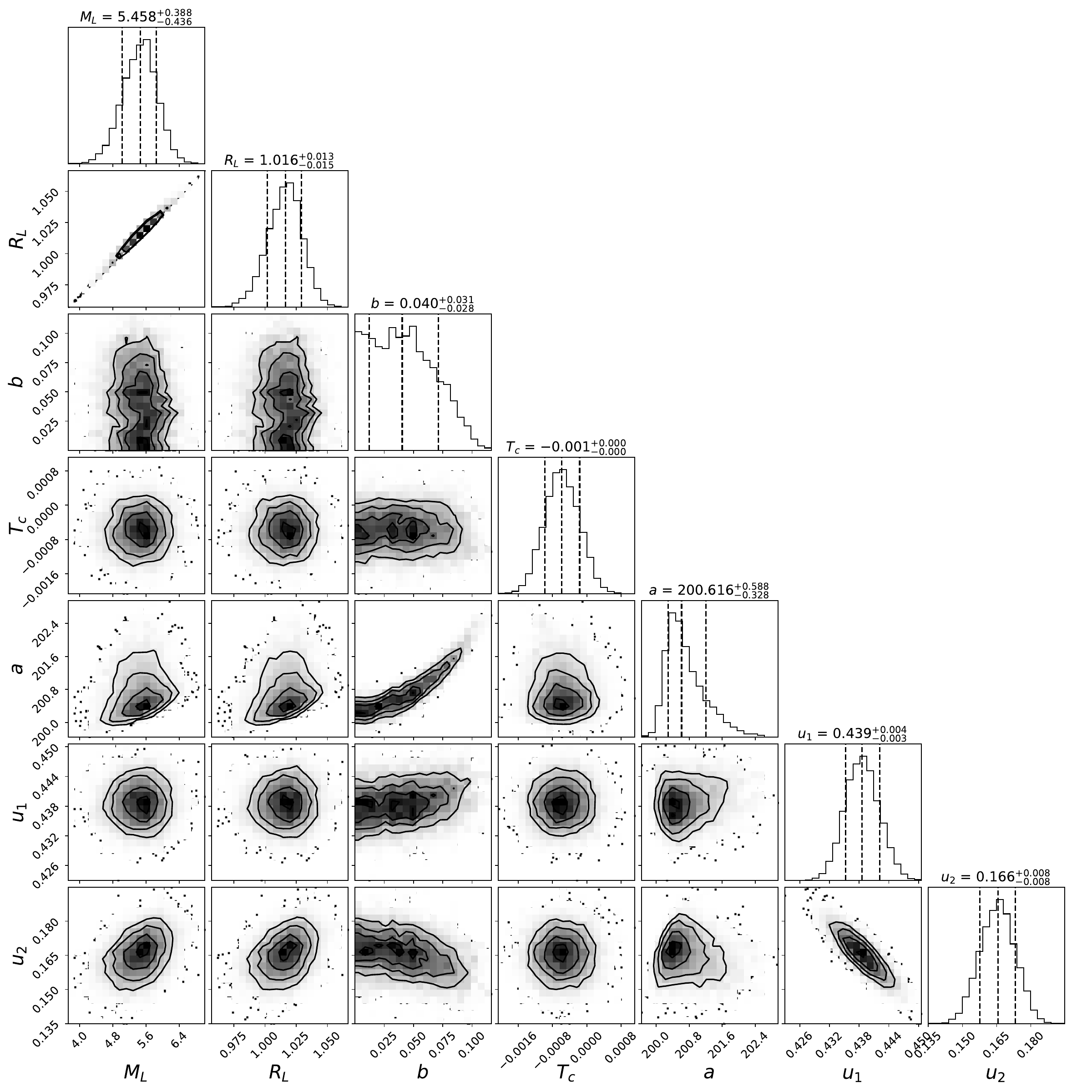}
\caption{Posterior distribution for the mock data with $a=100$~AU (left) and 
$a=200$~AU (right) for $M_L=5M_J$ including lensing effects.
\label{fig:mock_5_100-200}}
\end{figure}

Here we show the posterior distributions of full parameters for all the cases 
of the mock data in Sect.~\ref{sec:Mock} in Figures 
\ref{fig:mock_10_10-20}-\ref{fig:mock_5_100-200}.

\vspace*{5mm}

\vspace*{15mm}

\vspace*{15mm}

\section{Posterior distributions of parameters for the mock data in sect.5}
\label{app:Mock_R}
\begin{figure}[ht!]
\plottwo{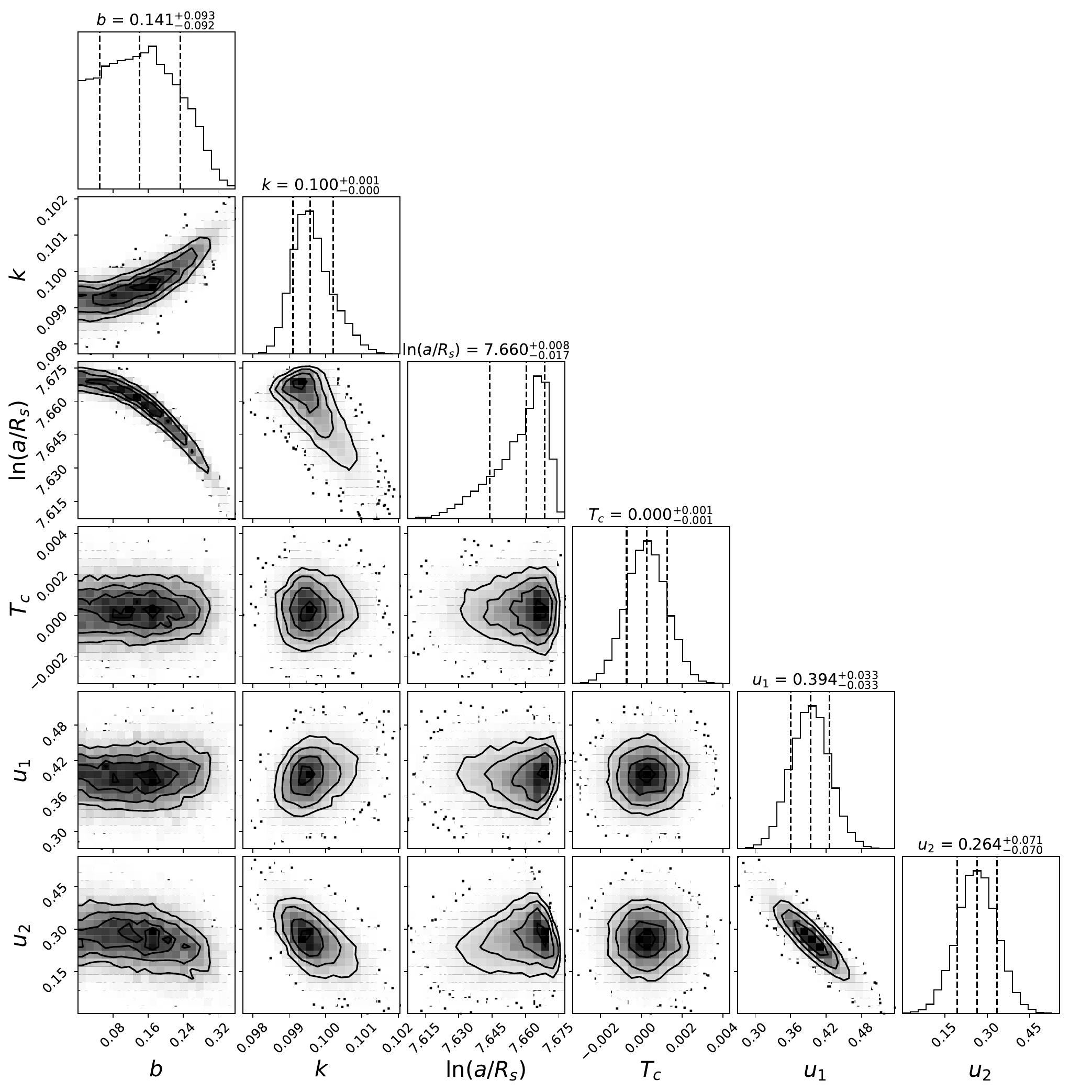}{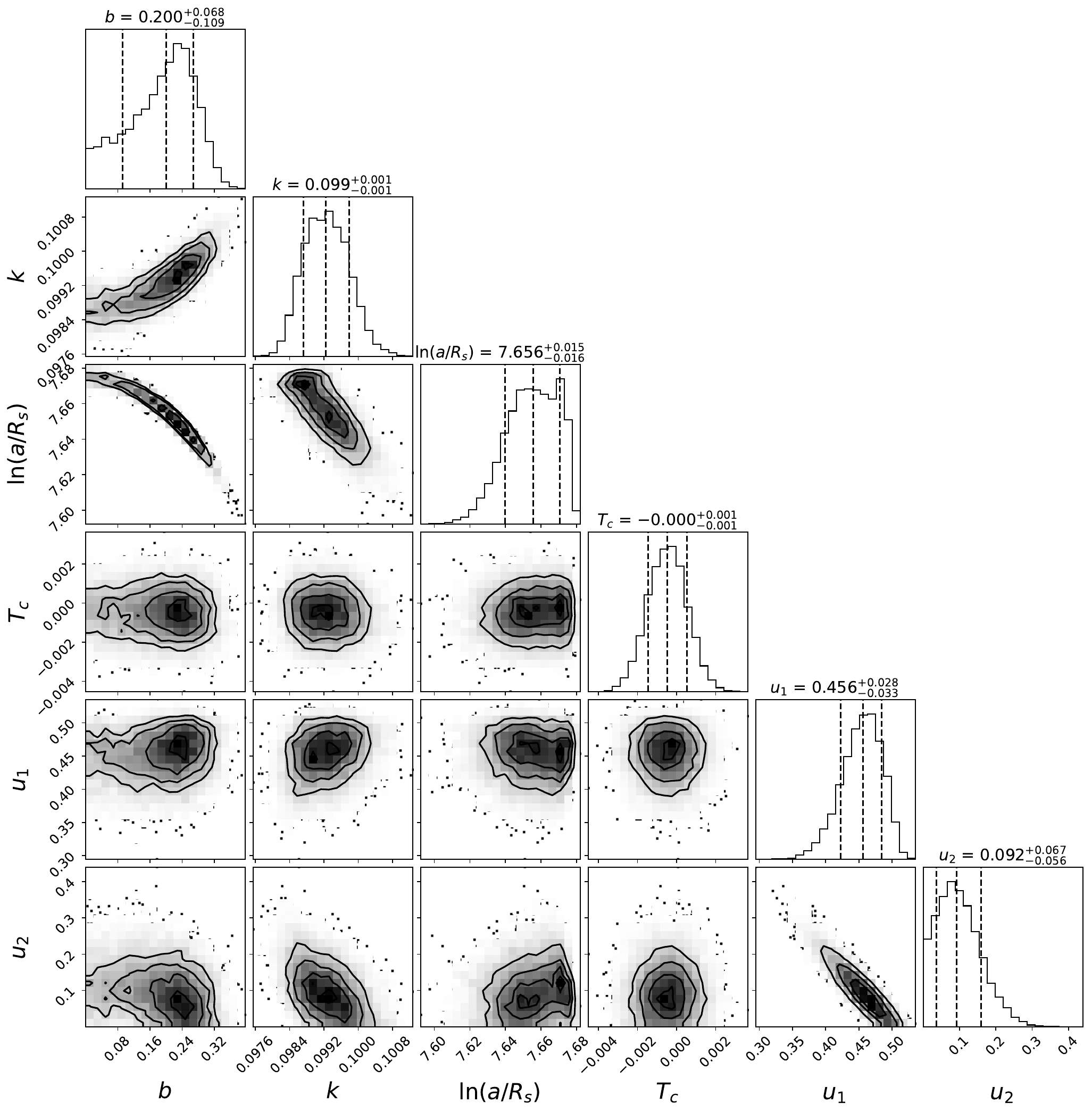}
\caption{Posterior distribution for the mock data for $M_L=5M_J$ (left) and
$M_L=10M_J$ (right) with $a=10$~AU.
\label{fig:mock_5_10_10}}
\end{figure}

\begin{figure}[ht!]
\plottwo{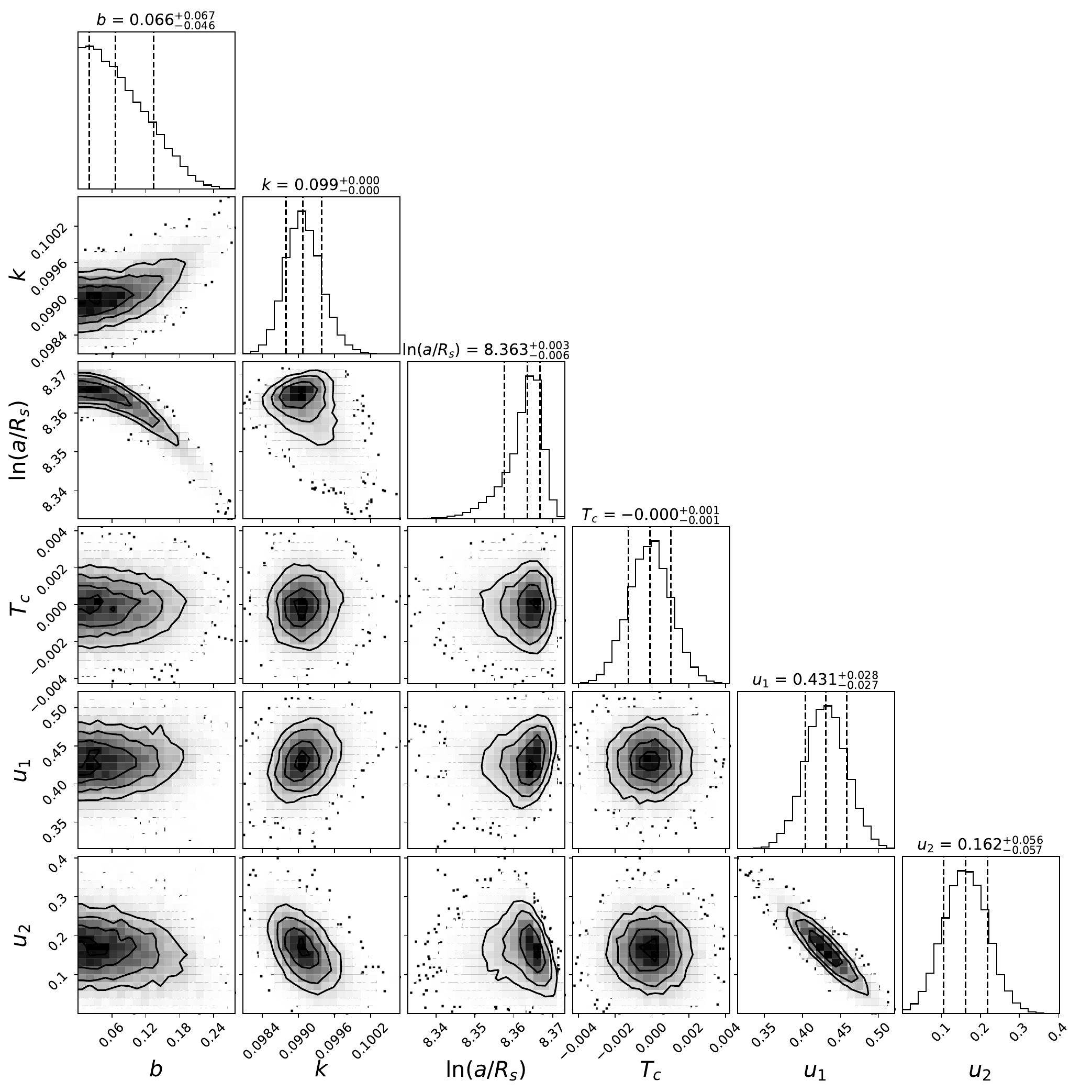}{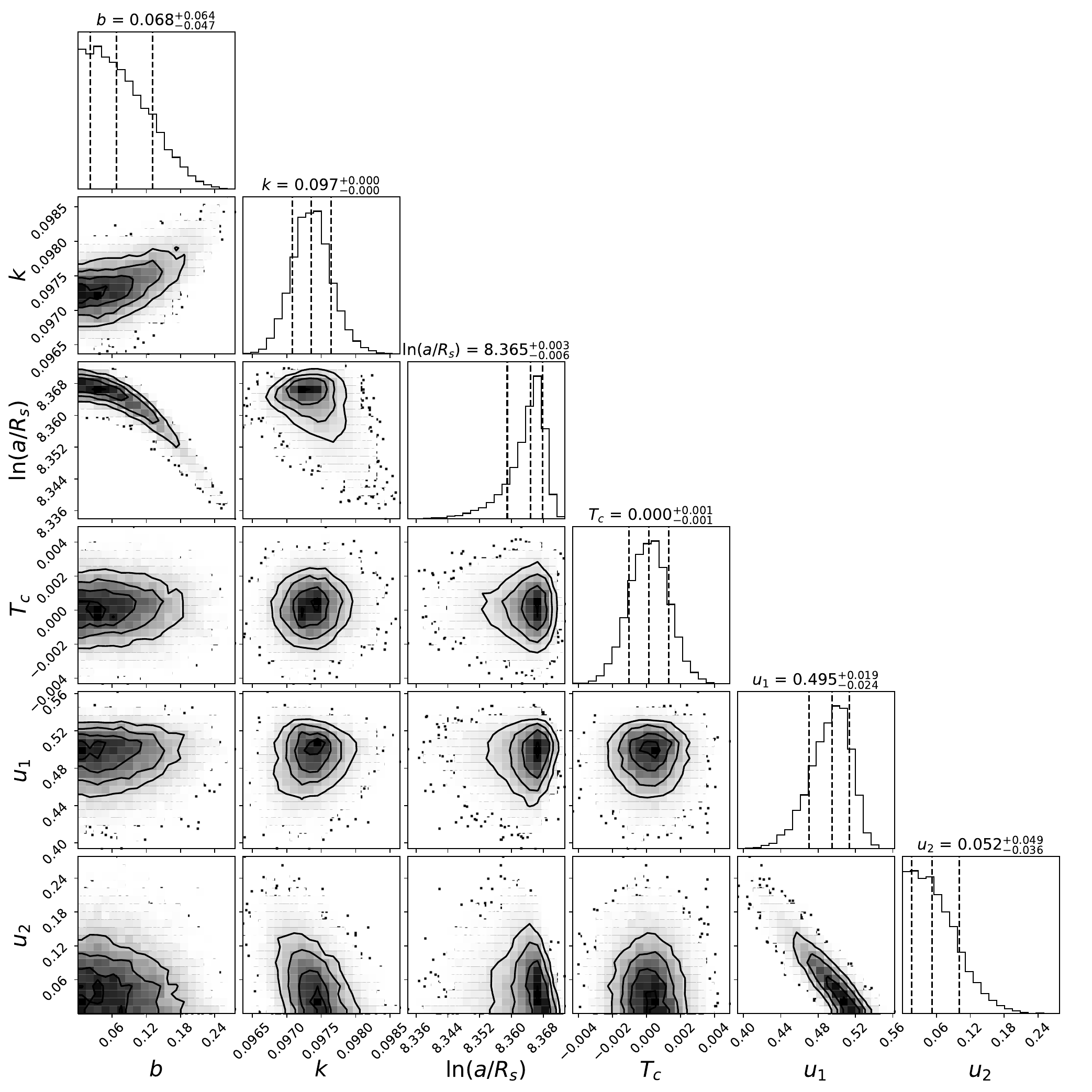}
\caption{Posterior distribution for the mock data for $M_L=5M_J$ (left) and
$M_L=10M_J$ (right) with $a=20$~AU.
\label{fig:mock_5_10_20}}
\end{figure}

\begin{figure}[ht!]
\plottwo{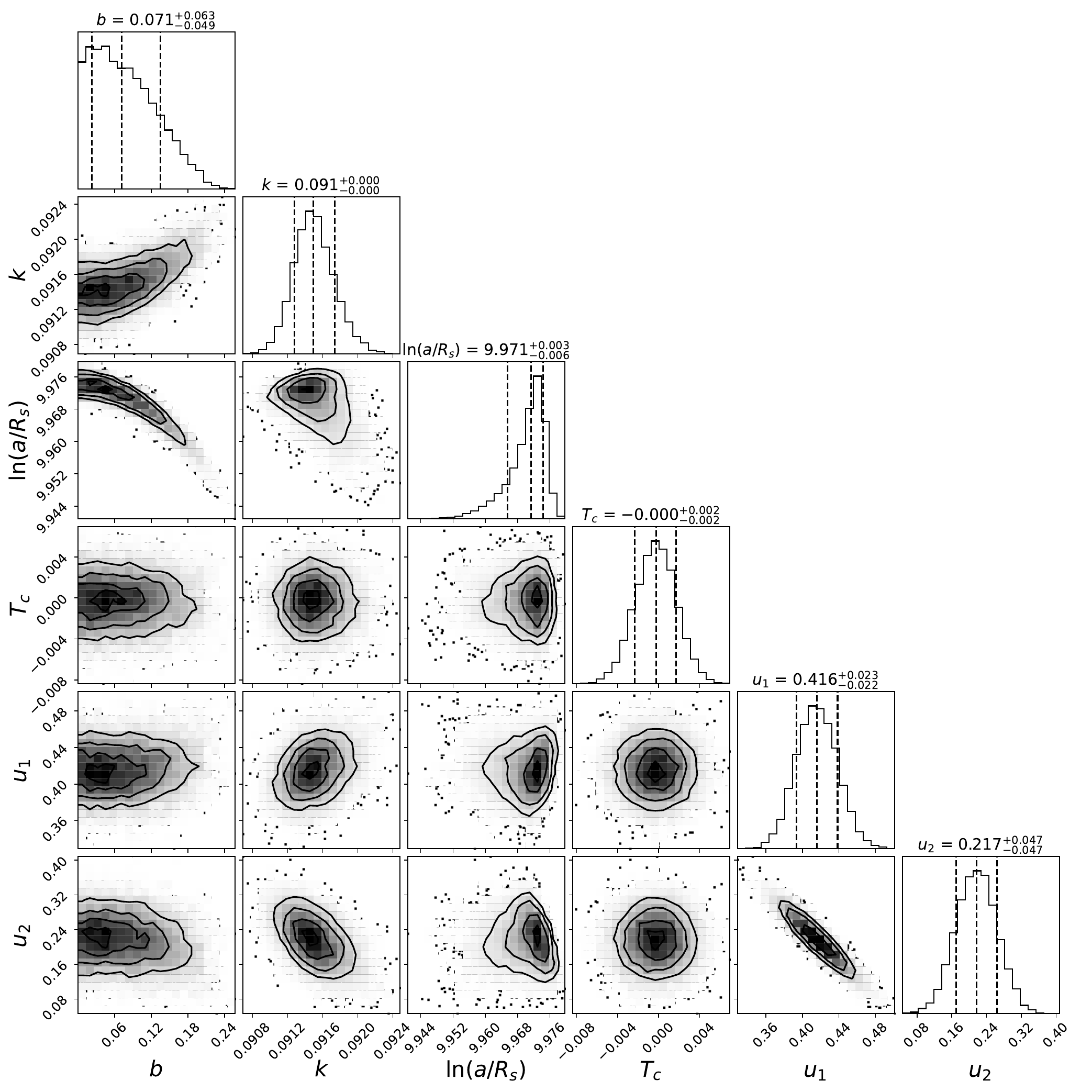}{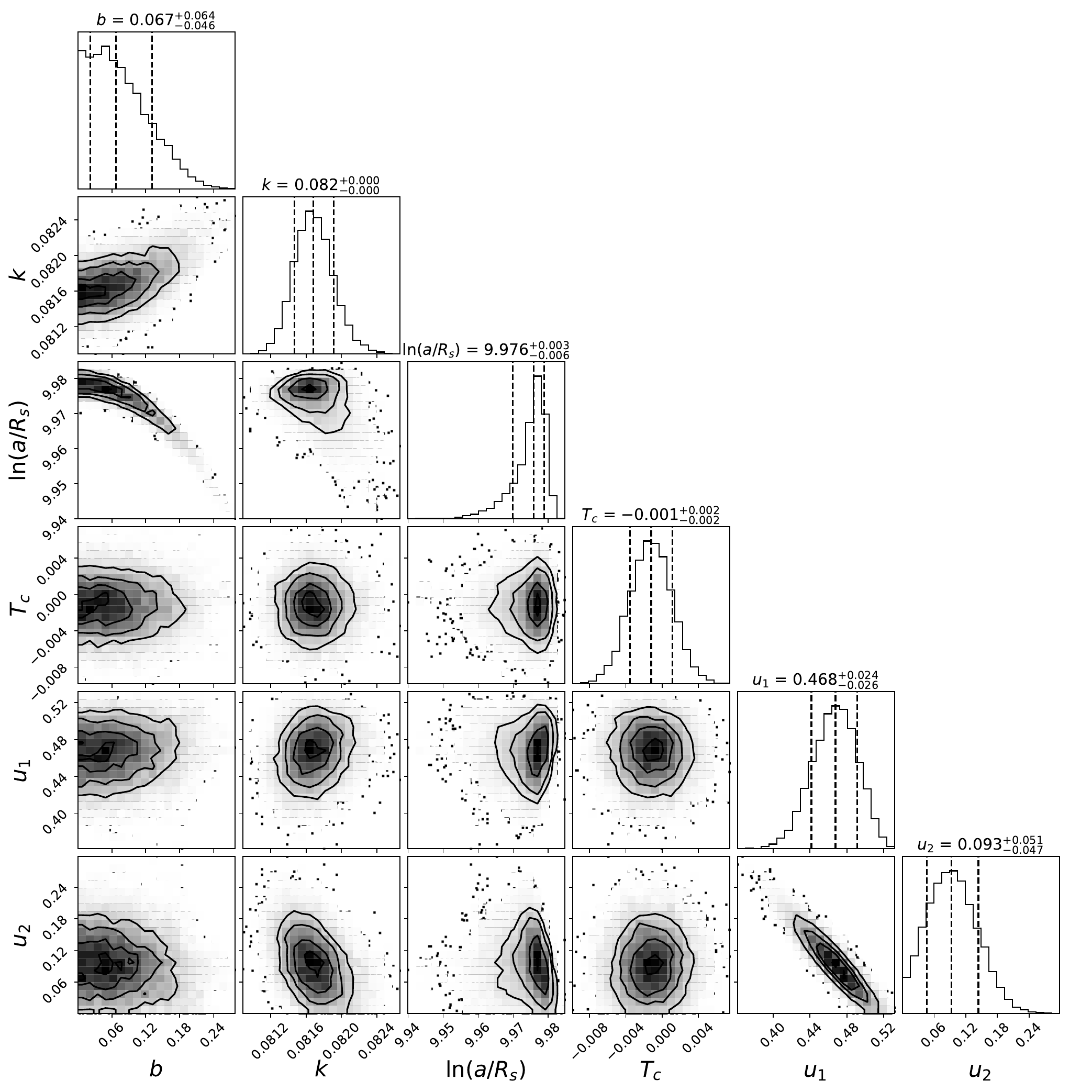}
\caption{Posterior distribution for the mock data for $M_L=5M_J$ (left) and
$M_L=10M_J$ (right) with $a=100$~AU.
\label{fig:mock_5_10_100}}
\end{figure}

We display below the posterior distributions of full parameters for all the cases 
of the mock data in Sect.~\ref{sec:radius} in Figures 
\ref{fig:mock_5_10_10}-\ref{fig:mock_5_10_100}.

\vspace*{15mm}

\vspace*{10mm}

\vspace*{10mm}
